\documentclass[twocolumn,5p,12pt,authoryear]{elsarticle}

\usepackage[switch]{lineno}

\newcommand*\patchAmsMathEnvironmentForLineno[1]{%
  \expandafter\let\csname old#1\expandafter\endcsname\csname #1\endcsname
  \expandafter\let\csname oldend#1\expandafter\endcsname\csname end#1\endcsname
  \renewenvironment{#1}%
     {\linenomath\csname old#1\endcsname}%
     {\csname oldend#1\endcsname\endlinenomath}}%
\newcommand*\patchBothAmsMathEnvironmentsForLineno[1]{%
  \patchAmsMathEnvironmentForLineno{#1}%
  \patchAmsMathEnvironmentForLineno{#1*}}%
\AtBeginDocument{%
\patchBothAmsMathEnvironmentsForLineno{equation}%
\patchBothAmsMathEnvironmentsForLineno{align}%
\patchBothAmsMathEnvironmentsForLineno{flalign}%
\patchBothAmsMathEnvironmentsForLineno{alignat}%
\patchBothAmsMathEnvironmentsForLineno{gather}%
\patchBothAmsMathEnvironmentsForLineno{multline}%
}

\usepackage[defaultcolor=red]{changes}

\usepackage{amsmath, amssymb, amsfonts, amscd}
\usepackage{natbib}
\usepackage[colorlinks,citecolor=magenta,linkcolor=blue]{hyperref}
\journal{Earth and Planetary Science Letters}

\def\time{t}

\newcommand{\RVA}[1]{{#1}}

\newcommand\RVR[2]{{#2}}

\begin{document}

\begin{frontmatter}


\title{Thermal tides in neutrally stratified atmospheres: Revisiting the Earth's Precambrian rotational equilibrium}

\author[1]{Mohammad Farhat\corref{cor1}}
\ead{mohammad.farhat@obspm.fr}
\author[1]{Pierre Auclair-Desrotour}
\author[1]{Gwena\"{e}l Bou\'{e}}
\author[2]{Russell Deitrick}

\author[1]{Jacques Laskar}
\affiliation[1]{organization={IMCCE, CNRS, Observatoire de Paris, PSL University, Sorbonne Université},
            addressline={77 Avenue Denfert-Rochereau}, 
            city={Paris},
            postcode={75014}, 
            country={France}}
\affiliation[2]{organization={School of Earth and Ocean Sciences, University of Victoria}, 
 city={ Victoria, British Columbia},
  country={Canada}
}
\cortext[cor1]{Corresponding author}

\begin{abstract}
Rotational dynamics of the Earth, over geological timescales, have profoundly affected local and global climatic evolution, probably contributing to the evolution of life. To better retrieve the Earth's rotational history, and motivated by the published hypothesis of a stabilized length of day during the Precambrian, we examine the effect of thermal tides on the evolution of planetary rotational motion. The hypothesized scenario is contingent upon encountering a \RVR{Lamb resonance}{resonance in atmospheric Lamb waves}, whereby an amplified thermotidal torque cancels the opposing torque of the oceans and solid interior, driving the Earth into a rotational equilibrium. With this scenario in mind, we construct an ab-initio model of thermal tides on rocky planets describing a neutrally stratified atmosphere. The model takes into account dissipative processes with Newtonian cooling and diffusive processes in the planetary boundary layer. We retrieve from this model a closed form solution for the frequency-dependent tidal torque which captures the main spectral features previously computed using 3D general circulation models. In particular, under longwave heating, diffusive processes near the surface and the delayed thermal response of the ground prove to be responsible for attenuating, and possibly annihilating, the accelerating effect of the thermotidal torque at the resonance. When applied to the Earth, our model prediction suggests the occurrence of the Lamb resonance in the Phanerozoic, but with an amplitude that is insufficient for the rotational equilibrium. Interestingly, though our study was motivated by the Earth's history, the generic tidal solution can be straightforwardly and efficiently applied in exoplanetary settings.

\end{abstract}



\end{frontmatter}



\section{Introduction}
For present day Earth, the semi-diurnal atmospheric tide, driven by the thermal forcing of the Sun and generated via atmospheric pressure waves, describes the movement of atmospheric mass away from the substellar point. Consequently, mass culminates forming bulges on the nightside and the dayside, generating a torque that accelerates the Earth's rotation. As such, this thermally generated torque counteracts the luni-solar gravitational torque associated with the Earth's solid and oceanic tides. The latter components typically drive the closed system of the tidal players towards equilibrium states of orbital circularity, coplanarity, and synchronous rotation via dissipative mechanisms \citep[e.g.,][]{mignard1980evolution,hut1981tidal}. In contrast, the inclusion of the stellar flux as an external source of energy renders the system an open system where radiative energy is converted, by the atmosphere, into mechanical deformation and gravitational potential energy. Though this competition between the torques is established on Earth, the thermotidal torque remains, at least currently, orders of magnitude smaller.

Interestingly though, this dominance of the gravitational torque over the thermal counterpart admits exceptions. The question of the potential amplification of the atmospheric tidal response initiated with \citet{thomson18822}, who invoked the theory of atmospheric tidal resonances, ushering a stream of theoretical studies investigating the normal modes spectrum of the Earth's atmosphere \citep[see][for a neat and authoritative historical overview]{chapman1969atmospheric}. Studies of the Earth's tidal response spectrum 
advanced the theory of thermal tides for it to be applied to Venus \citep{goldreich1966q,gold1969atmospheric,ingersoll1978venus,dobrovolskis1980atmospheric,correia2001four,correia2003long2,correia2003long}, hot Jupiters \citep[e.g.,][]{arras2010thermal,auclair2018semidiurnal,gu2019modeling,lee2020tidal}, and near-synchronous and Earth-like rocky exoplanets \citep{cunha2015spin,leconte2015asynchronous,auclair2017atmospheric,auclair2019generic}. Namely, for planetary systems within the so-called  habitable zone, the gravitational tidal torque diminishes in the regime near spin-orbit synchronization and becomes  comparable in magnitude to the thermotidal torque. Consequently, the latter may actually prevent the planet from  precisely reaching its destined synchronous state \citep{laskar2004rotation,correia_tidal_2010,cunha2015spin,leconte2015asynchronous}.



\begin{figure}[h]
\centering
\includegraphics[width=.47\textwidth]{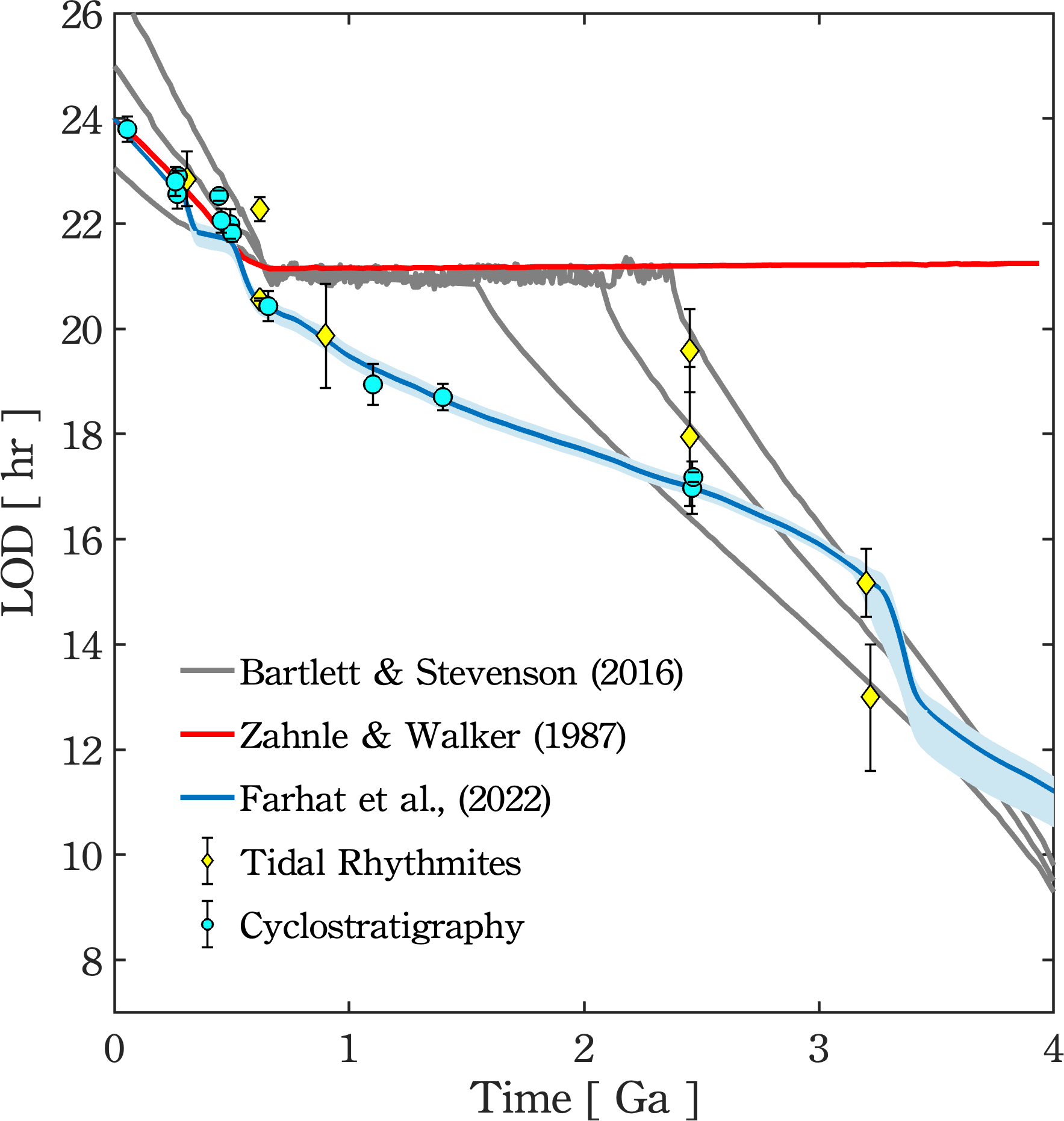}
    \caption{Modeled histories of the rotational motion of the Earth. Plotted is the Earth's LOD evolution in time over geological timescales for three models: \textit{i)} the model of \cite{farhat2022resonant}, where the evolution is driven solely by oceanic and solid tidal dissipation; \textit{ii)} the model of \cite{zahnle1987constant}, where the Lamb resonance is encountered for LOD${\sim}21$  hr, forcing a rotational equilibrium on the Earth; \textit{iii)} the model of \cite{bartlett2016analysis}, which also adopts the equilibrium scenario, but further studies the effect of thermal noise, and the required temperature variation to escape the equilibrium. Three curves of the latter model correspond to different  parameterizations of the gravitational tide. Plotted on top of the modelled  histories are geological proxies of the LOD evolution that can be retrieved from \url{http://astrogeo.eu}.}
    \label{LOD_evo_models}
\end{figure}

Going back to Earth, \cite{holmberg1952suggested} suggested that the thermal tide at present is resonant, and the generated torque is  equal in magnitude and opposite in sign to that generated by gravitational tides, thus placing the Earth into a rotational equilibrium with a stabilized spin rate. As this was proven to be untrue for present Earth \citep{chapman1969atmospheric}, \cite{zahnle1987constant} revived Holmberg's hypothesis by applying the resonance scenario of thermal tides to the distant past. Their suggestion relied on two factors needed to close the gap between the competing torques. The first is the occurrence of a resonance in atmospheric Lamb waves \citep[e.g.,][]{lindzen1972lamb} -- which we coin as a Lamb resonance -- that characterizes the frequency overlap between the fundamental mode of atmospheric free oscillations and the semidiurnal forcing frequency. According to \cite{zahnle1987constant}, this resonance occurred when the length of day (LOD) was around 21 hrs, exciting the thermotidal torque to large amplitudes. Secondly, the gravitational tidal torque must have been largely attenuated in the Precambrian.
Recently, \cite{bartlett2016analysis} revisited the equilibrium scenario and investigated the effect of temperature fluctuations on the stability of the resonance trapping and the Earth's  equilibrium. The authors concluded that the rotational stabilization could have lasted 1 billion years, only to be distorted by a drastic deglaciation event (on the scale that follows the termination of a snowball Earth), thus allowing the LOD to increase again from ${\sim}21$ hr to its present value. Evidently, the occurrence of such a scenario has very significant implications on paleoclimatic studies, with the growing evidence on links between the evolving LOD and the evolution of Precambrian benthic life \citep[e.g.,][]{klatt2021possible}.

We are fresh out of a study on the tidal evolution of the Earth-Moon system \citep{farhat2022resonant}, where we focused on modelling tidal dissipation in the Earth's paleo-oceans and solid interior. There we learned that the tidal response of the oceans, characterized by intermittent resonant excitations, is sufficient to explain the present rate of lunar recession and the estimated lunar age, and is in good agreement with the geological proxies on the lunar distance and the LOD, leaving little-to-no place for an interval of a rotational equilibrium (Figure \ref{LOD_evo_models}). 

On the other hand, major progress has been achieved in establishing the frequency spectrum of the thermotidal response of rocky planets with various approaches ranging from  analytical models \citep{ingersoll1978venus,dobrovolskis1980atmospheric,auclair2017atmospheric,auclair2017rotation}, to parameterized models that capture essential spectral features \citep[e.g.,][]{correia2001four,correia2003different}, to fully numerical efforts that relied on the advancing sophistication of general circulation models \citep[GCM; e.g.,][]{leconte2015asynchronous,auclair2019generic}. The latter work presents, to our knowledge, the first and, to-date\footnote{\RVA{While this paper was under review, \cite{wu2023day} presented another GCM-computed spectrum for the Earth in the high frequency regime. We provide an elaborate discussion of their work's results in a separate dedicated paper \citep{Laskar23}.}}, the only study to have numerically computed the planetary thermotidal torque in the high frequency regime, i.e. around the Lamb resonance \citep{lindzen1972lamb}. Of interest to us here are two perplexing results that \citet{auclair2019generic} established: first, for planets near synchronization, the simplified Maxwellian models often used to characterize the thermotidal torque did not match the GCM simulated response; second, the torque at the Lamb resonance featured only a decelerating effect on the planet. Namely, it acts in the same direction of gravitational tides, and thus the effect required for the rotational stabilization disappeared.

\RVA{More recently, while this work was under review, two studies on the Precambrian LOD stabilization were  published. \cite{mitchell2023mid} compiled various geological proxies on the Precambrian LOD and established the best piece-wise linear fit to this data compilation. The authors' analysis depicts that a Precambrian LOD of 19 hr was stabilized between 1 and 2 Ga. In parallel, \cite{wu2023day} also attempted to fit a fairly similar set of geological proxies, but using a simplified model of thermal tides. The authors conclude that the LOD was stabilized at $\sim19.5$~hr between 0.6 and 2.2 Ga, with a sustained very high mean surface temperature ($40-55^\circ$C). Although using different approaches, the two studies have thus arrived at similar conclusions. A closer look at the subset of geological data that favored this outcome, however, indicates that both studies heavily rely on three stromatolitic records from the Paleoproterozoic that were originally studied by \cite{pannella1972paleontological,pannella1972precambrian}. These geological data have been, ever since, identified as unsuitable for precise quantitative interpretation \citep[see e.g.,][]{scrutton1978periodic,lambeck2005earth,williams2000geological}.
To this end, we provide a more detailed analysis of the geological proxies of the LOD, and of the model presented by \cite{wu2023day} in a parallel paper dedicated to the matter \citep{Laskar23}.}
   
With \RVA{a} view to greater physical realism, we aim here to study, analytically, the frequency spectrum of the thermotidal torque, from first principles, interpolating between the low and high frequency regimes. Our motivation is two-fold: first, to provide a novel physical model for the planetary thermotidal torque that better matches the GCM-computed response, and that can be used in planetary dynamical evolution studies; second, to apply this model to the Earth and attempt quantifying the amplitude of the torque at the Lamb resonance and explore the intriguing  rotational equilibrium scenario. 

\section{Ab initio atmospheric dynamics}
For an atmosphere enveloping a spherically symmetric planet, we define a reference frame co-rotating with the planet. In this frame, an atmospheric parcel is traced by its position vector $\boldsymbol{r}$ in spherical coordinates ($r,\theta,\varphi)$, such that $\theta$ is the colatitude, $\varphi$ is the longitude, and the radial distance $|\boldsymbol{r}|=R_{\rm p} + z$, where $R_{\rm p}$ is the planet's radius and $z$ is the parcel's atmospheric altitude. The atmosphere is characterized by the scalar fields of pressure $p$, temperature $T$, density $\rho$, and the three-dimensional vectorial velocity field $\boldsymbol{\mathcal{V}}$. Each of these fields varies in time and space, and is decomposed linearly into two terms: a background, equilibrium state field, subscripted with $0$, and a tidally forced perturbation term of significantly smaller amplitude such that
$p=p_0 + \delta p, \, T=T_0 + \delta T,\,  \rho=\rho_0 + \delta\rho,\,$and\,$\boldsymbol{\mathcal{V}}=\boldsymbol{V}_0 + \boldsymbol{V}.$ 

Our fiducial atmosphere is subject to the perturbative gravitational tidal potential $U$ and the thermal forcing per unit mass $J$. We shall define the latter component precisely in Section \ref{Sec_thermal_forcing}, but for now it suffices to say that $J$ accounts for the net amount of heat, per unit mass, provided to the atmosphere, allowing for thermal losses driven by radiative dissipation. We take the latter effect into account by following the linear Newtonian cooling hypothesis\footnote{\RVA{noting that surface friction is another dissipative mechanism as discussed by \cite{lindzen1972lamb}.}} \citep{lindzen1967tidal},  where radiative losses, $J_{\rm rad}$, are  parameterized by the characteristic frequency $\sigma_0$;  namely $J_{\rm rad}= p_0\sigma_0/(\kappa\rho_0T_0) \delta T$, where $\kappa=(\Gamma_1-1)/\Gamma_1=0.285$ and $\Gamma_1$ is the adiabatic exponent. \RVA{Similar to \cite{leconte2015asynchronous}, we associate with $\sigma_0$ a radiative cooling timescale $\tau_{\rm rad}=4\pi/\sigma_0.$}
\subsection{The vertical structure of tidal dynamics}
We are interested in providing a closed form solution for the frequency\footnote{The frequency in this case being the tidal forcing frequency $\sigma$, typically a linear function of the planet's spin rate $\Omega$ and the stellar perturber's mean motion $n_\star$. The semi-diurnal tidal frequency, for instance, $\sigma_{22}=2(\Omega-n_\star).$} dependence of the thermotidal torque, which results from tidally driven atmospheric mass redistribution. By virtue of the hydrostatic approximation, this mass redistribution is encoded in the vertical profile of pressure. As such, it is required to solve for the vertical structure of tidal dynamics. With fellow non-theoreticians in mind, we delegate the detailed development of the governing system of equations describing the tidal response of the atmosphere to \ref{Appendix_non_dimensionization}. Therein, we employ the classical system of primitive equations describing momentum and mass conservation \citep[e.g.,][]{siebert1961atmospheric,chapman1969atmospheric}, atmospheric heat transfer augmented with linear radiative transfer à la \cite{lindzen1967tidal}, and the ideal gas law, all \RVR{neatly}{} formulated in a dimensionless form. 

Aided by the so-called traditional approximation \citep[e.g.,][see also \ref{Appendix_non_dimensionization}]{unno1989nonradial}, the analytical treatment of the said system is feasible as it decomposes into two parts describing, separately, the horizontal and vertical structures of tidal flows.  The former part is completely described by the eigenvalue-eigenfunction problem defined as Laplace's tidal equation \citep{laplace1798traite,lee1997low}:
\begin{equation}\label{Laplace_Tidal_equation}
\mathcal{L}^{m,\nu}\Theta^{m,\nu}_n = -\Lambda^{m,\nu}_n\Theta^{m,\nu}_n,
\end{equation}
where the set of Hough functions $\{\Theta^{m,\nu}_n\}$  serves as the solution  \citep{hough}, $\{\Lambda^{m,\nu}_n\}$ is the associated set of eigenvalues, $\mathcal{L}^{m,\nu}$ is a horizontal operator defined in Eq.\eqref{horizontal_operator} of \ref{Appendix_non_dimensionization}, while $\nu=2\Omega/\sigma$, where $\Omega$ is the rotational velocity of the planet and $\sigma$ is the tidal forcing frequency. In the tidal system under study,  the variables and functions $(\delta p, \delta \rho,\delta T, \boldsymbol{V}, J, \Theta,\Lambda)$ are written in the Fourier domain using the longitudinal order $m$ and frequency $\sigma$ (Eq.\ref{Fourier_expansion}), and expanded in horizontal Hough modes with index $n$ (Eq.\ref{Hough_expansion}). We denote hereafter their  coefficients $f_n^{m,\nu}$ by $f_n$ to lighten the expressions. This horizontal structure of tidal dynamics is merely coupled to the vertical structure via the set of eigenvalues $\{\Lambda^{m,\nu}_n\}$. To construct these sets of eigenfunctions-eigenvalues we use the spectral method laid out by \cite{wang2016computation}. 

The vertical structure on the other hand requires a more elaborate manipulation of the governing system, a procedure that we detail in \ref{VSE_appendix}. The outcome is a wave-like equation that describes vertical thermotidal dynamics and reads as:
\begin{equation}\label{wave_equation}
    \frac{d^2\Psi_n}{dx^2} + \hat{k}_n^2\Psi_n = \Phi^{-1}C_n.
\end{equation}
Here, as is the common practice \citep[e.g.,][]{siebert1961atmospheric,chapman1969atmospheric}, we use the reduced altitude $x=\int_0^z dz/H(z)$ as the vertical coordinate, where the pressure scale height $H(z)=\mathcal{R}_{\rm s}T_0(z)/g$; $\mathcal{R}_{\rm s}$ being the specific gas constant and $g$ the gravitational acceleration. The quantity $\Psi_n(x)$ is a calculation variable from which, once solved for, all the tidal scalar and vectorial quantities would flesh out (\ref{polarizations_appendix}). The vertical wave number $ \hat{k}_n(x)$ is defined via
\begin{align}\nonumber
    \hat{k}_n^2(x) &=-\frac{1}{4}\Bigg\{\!\!\!\left(\!1\!-\!\frac{i\kappa}{\alpha-i}(\gamma-1)\right)^2\!\!\!\!+2\frac{d}{dx}\!\!\left(\frac{i\gamma}{\alpha-i}\right)\\
    &-\frac{4\alpha\kappa}{\alpha-i}\!\left[\beta\gamma\Lambda_n-\!\frac{i}{\alpha}(\gamma-1) \right]\Bigg\}.\label{wavenumber2}
\end{align}
\begin{table}[]
\centering
\caption{\label{tab:control_param} Dimensionless control parameters determining the regime of the atmospheric tidal response and defined throughout the text. }
\vspace{.3cm}
\begin{tabular}{ll} 
 \hline 
 \hline 
\textsc{Parameter} & \textsc{Scale significance of} \\
 \hline \\[-0.3cm] 
 $ \eta = {R_{\rm p}}/{H} $& horizontal structure of dynamics \\[0.3cm] 
$ \nu = {2 \Omega}/{\sigma} $& Coriolis effects  \\[0.3cm] 
$ \alpha = {\sigma}/{\sigma_0} $&  radiative cooling   \\[0.3cm] 
$ \beta = {\sigma_{\rm w}^2}/{\sigma^2} $&  Lamb waves  \\[0.3cm] 
$ \gamma = {N_{\rm B}^2}/{N_{\rm B; iso}^2} $ & buoyancy forces \\[0.3cm] 
$ \zeta = \sqrt{{|\sigma|}/\sigma_{\rm bl}} $& ground and atmospheric inertia\\[0.3cm] 
$ \mu_{\rm gr} = \frac{I_{\rm gr}}{I_{\rm gr}+I_{\rm atm}} $&  ground thermal inertia\\[0.3cm] 
\hline
 \end{tabular}
 \end{table}
 
By virtue of the non-dimensionalization of the governing system of equations, dimensionless control parameters appear in the wavenumber definition. Namely, $\alpha = \sigma/\sigma_0$, $\beta = {\sigma_{\rm w}^2}/{\sigma^2}$, and $\gamma={N_{\rm B}^2}/{N_{\rm B; iso}^2}$, where we have introduced the characteristic frequency $\sigma_{\rm w}=\sqrt{gH}/R_{\rm p}$, a typical frequency of Lamb waves.
Of significance to the computations that follow is the Brunt–Väisälä frequency, $N_{\rm B}$, a measure of the vertical density stratification of the atmosphere against the strength of convection \citep[e.g,][]{vallis2017atmospheric}.  Under hydrostatic equilibrium, $N_{\rm B}$ is defined as:
\begin{equation}\label{Brunt_Vaisala2}
     N_{\rm B}^2 =\frac{g}{H}\left(\kappa + \frac{d\ln H}{dx}\right).
\end{equation}
\RVA{We define the Brunt–Väisälä frequency in the limit of an isothermal atmospheric profile, $N_{\rm B; iso}=\sqrt{\kappa g/H}$. As such, the parameter $\gamma$ measures the local atmospheric stability against convection with respect to the stability of an equivalent isothermal atmosphere.} These key control parameters, among others, qualitatively determine the regime of the tidal response. We summarize these parameters in Table \ref{tab:control_param}. 
The right hand side of the wave equation \eqref{wave_equation} combines the function $\Phi$ defined as
\begin{equation}\label{Eq_Phi_x}
    \Phi(x)= \exp\left[\frac{x}{2} - \frac{1}{2}\int_0^x\frac{i}{\alpha-i}\left(\frac{HN_{\rm B}^2}{g}-\kappa\right)dx^\prime\right],
\end{equation}
and the forcing function $C_n(x)$ which reads:
\begin{align}\nonumber
    C_n (x)&= \frac{\alpha\beta\Lambda_n\tilde{J}_n}{\alpha-i} + \Bigg[\frac{\kappa(\gamma-1)}{\alpha-i}\left(\Lambda_n+\frac{\partial_x}{\beta}\right)\\\label{C_equation}
   & -i\partial_x\left(\frac{\partial_x}{\beta}\right)\!\Bigg]\tilde{U}_n.
\end{align}
Hereafter, we identify the dimensionless form of the tidal variables by the tilde diacritic. Namely in Eq.\eqref{C_equation}, the dimensionless thermal forcing function $\tilde{J}_n=\kappa/(g\sigma H)J_n$, while the gravitational analogue $\tilde{U}_n=U_n/v_0^2$, where the reference velocity $v_0=\sigma R_{\rm p}$ (see Eq.\eqref{non_dimensionlaizaion_eqns} for the rest of the dimensionless quantities). As we intend to solve the wave equation in the following sections, what is left for us to quantify the mass redistribution and compute the resulting tidal torque is to retrieve the vertical profile of pressure given the solution of the wave equation, $\Psi_n(x)$. In \ref{polarizations_appendix}, we derive the vertical profiles of all the tidal variables, and specifically for the dimensionless pressure anomaly we obtain:
\begin{equation}\label{pressure_anomaly}
    \tilde{\delta p}_n (x)= \frac{1}{i\beta\Lambda_n}\left(\frac{d\tilde{G}_n}{dx}-\tilde{G}_n\right) + \frac{1}{\beta}\left(1+\frac{1}{\beta\Lambda_n}\frac{d}{dx}\right)\tilde{U}_n,
\end{equation}
where $\tilde{\delta p}_n(x)=\delta p_n(x)/p_0$, and the calculation variable $\tilde{G}_n(x)=\Psi_n(x)\Phi(x)$.
\subsection{The thermal forcing profile}\label{Sec_thermal_forcing}
To solve the non-homogeneous wave equation \eqref{wave_equation}, it is necessary to define a vertical profile for the tidal heating power per unit mass $\tilde{J}_n$ (or equivalently in dimensional form, $J_n$). We adopt a vertical tidal heating profile of the form
\begin{equation}\label{thermal_profile}
    J_n(x) = J_{\rm s} e^{-b_{\rm J}x},
\end{equation}
where $J_{\rm s}$ is the heat absorbed at the surface and $b_{\rm J}$ is a decay rate that characterizes the exponential decay of heating along the vertical coordinate. As we are after a generic planetary model, this functional form of $J_n$ allows the distribution of heat  to vary between the Dirac distribution adopted by 
\cite{dobrovolskis1980atmospheric} where $b_{\rm J}\rightarrow\infty$, and a uniform distribution where the whole air column is uniformly heated ($b_{\rm J}=0)$. 

To determine $J_{\rm s}$, we invoke its dependence on the total vertically propagating flux $\delta F_{\rm tot}$ by computing the energy budget over the air column. The net input of energy corresponds to the difference between the amount of flux absorbed by the column and associated with a local increase of thermal energy, and the amount that escapes into space or into the mean flows defining the background profiles. We quantify the fraction of energy transferred to the atmosphere and that is available for tidal dynamics by 
$\alpha_{\rm A}$, where $0\leq\alpha_{\rm A}\leq1$; the rest of the flux amounting to $1-\alpha_{\rm A}$ escapes the thermotidal interplay. We thus have
\begin{equation}\label{Jx_integral}
\int_0^\infty J(x)\rho_0(x)H(x) dx = \alpha_{\rm A}\delta F_{\rm tot}.
\end{equation}
To define $\delta F_{\rm tot}$, we establish the flux budget for a small thermal perturbation at the planetary surface. We start with $\delta F_{\rm inc}$, a variation of the effective incident stellar flux, after the reflected  component  has  been  removed. $\delta F_{\rm inc}$ generates a variation $\delta T_{\rm s}$ in the surface temperature $T_{\rm s}$. The proportionality between $\delta F_{\rm inc}$ and $\delta T_{\rm s}$ is parameterized by  $\tau_{\rm bl}$, a characteristic diffusion timescale of the ground and atmospheric surface thermal responses. We detail on this proportionality in \ref{appendix_thermal_Pbl}, but for now it suffices to state that $\tau_{\rm bl}$ is a function of the thermal inertia budgets in the ground, $I_{\rm gr}$, and the atmosphere $I_{\rm atm}$. We associate with $\tau_{\rm bl}$ the frequency $\sigma_{\rm bl}=\tau_{\rm bl}^{-1}$, a characteristic frequency that reflects the thermal properties of the diffusive boundary layer. It will serve as another free parameter of our tidal model, besides the Newtonian cooling frequency $\sigma_0$, and the atmospheric opacity parameter $\alpha_{\rm A}$. In analogy to $\alpha=\sigma/\sigma_0$, we define the dimensionless parameter for the boundary layer $\zeta=\sqrt{|\sigma|\tau_{\rm bl}}=\sqrt{|\sigma|/\sigma_{\rm bl}}$.

 
By virtue of the power budget balance established  in \ref{appendix_thermal_Pbl}, we define the total propagating flux $\delta F_{\rm tot}$ as
\begin{equation}\label{delta_f_tot}
    \delta F_{\rm tot} = \delta F_{\rm inc}\left[ 1- \mu_{\rm gr}\zeta\frac{1+si}{1+(1+si)\zeta}  \right].
\end{equation}
Here, $s={\rm sign}(\sigma)$, and $\mu_{\rm gr}$ is a dimensionless characteristic function weighing the relative contribution of ground thermal inertia to the total inertia budget; namely $ \mu_{\rm gr} = {I_{\rm gr}}/({I_{\rm gr}+I_{\rm atm}})$.

The generic form of the flux in Eq.(\ref{delta_f_tot}) clearly depicts two asymptotic regimes of thermotidal forcing:
\begin{itemize}
    \item[\textit{i)}]  Ignoring the surface layer effects associated with the term on the right, i.e. setting $\zeta=\mu_{\rm gr}=0$, leaves us with thermotidal heating that is purely attributed to the direct atmospheric absorption of the incident flux. This limit can be used to describe the present understanding of thermotidal forcing on Earth where, to first order, direct insolation absorption in the shortwave by ozone and water vapor appears sufficient to explain the observed tidal amplitudes in barometric measurements \citep[e.g.,][]{chapman1969atmospheric,schindelegger2014surface}. Nevertheless, it is noteworthy that the observed tidal phases of pressure maxima could not be explained by this direct absorption, a discrepancy later attributed to an additional semidiurnal forcing, namely that of latent heat release associated with cloud and raindrop formation \citep[e.g.,][]{lindzen1978effect,hagan2002migrating,sakazaki2017there}.
    \item[\textit{ii)}] Allowing for the surface layer term on the other hand ($\zeta\neq0$, $\mu_{\rm gr}\neq0$) places us in the limit where the ground radiation in the infrared and heat exchange processes occurring in the vicinity of the surface would dominate the thermotidal heating. The total tidal forcing in this case is non-synchronous with the incident flux due to the delayed thermal response of the ground, which here is a function of $\tau_{\rm bl}$. This limit better describes dry Venus-like planets, as is the fiducial setting studied using GCMs in \cite{leconte2015asynchronous} and \cite{auclair2019generic}.
\end{itemize}
Finally, as we are interested in the semi-diurnal tidal response, we decompose the thermal forcing in \ref{quadrupolar_forcing_appendix} to obtain the amplitude of the quadrupolar component as $ \delta F_{\rm inc} =\delta F_{22}=({\sqrt{30\pi}}/{16})F_{\star}$\,, where $F_{\star}= L_{\star}/4\pi a_{\rm p}^2$, $ L_{\star}$ being the stellar luminosity, and $a_{\rm p}$ the star-planet distance. 

\section{The tidal response}
\subsection{The tidal torque in the neutral stratification limit}\label{Section_Response_NS}
Under the defined forcing in the previous section, to solve the wave equation analytically, a choice has to be made on the Brunt–Väisälä frequency, $N_{\rm B}$ (Eq.\ref{Brunt_Vaisala2}), which describes the strength of atmospheric buoyancy forces and consequently the resulting vertical temperature profile. Earlier analytical solutions have been obtained in the limit of an isothermal atmosphere \citep{lindzen1967tidal,auclair2019generic}, in which case the scale height $H$ becomes independent of the vertical coordinate, and by virtue of Eq. \eqref{Brunt_Vaisala2}, $N_{\rm B}^2=\kappa g/H={\rm const}.$

Motivated by the Earth's atmosphere, where the massive troposphere (${\sim}80\%$ of atmospheric mass) controls the tidal mass redistribution, we derive next an analytical solution in a different, and perhaps more realistic limit. Namely, the limit corresponding to the case of a neutrally stratified atmosphere, where $N_{\rm B}^2=0.$ In fact, $N_{\rm B}^2$ can be expressed in terms of the potential temperature $\Theta_0$ \citep[e.g., Section 2.10 of][]{vallis2017atmospheric}:
\begin{equation}
    N_{\rm B}^2 = \frac{g}{H}\frac{d\ln\Theta_0}{dx}.
\end{equation}
whereby the stability of the atmosphere is controlled by the slope of $\Theta_0$. That said, atmospheric temperature measurements \citep[e.g., Figures 2.1-2.3 of][]{pierrehumbert2010principles} clearly depict that the troposphere is characterized by a negative temperature gradient, and a very weak potential temperature gradient, which is closer to an idealised adiabatic profile than it is to an idealised isothermal profile. Moreover, the heating in the troposphere generates strong convection and efficient turbulent stirring, thus enhancing energy transfer and driving the layer towards an adiabatic temperature profile. As such, the temperature profile being adiabatic would prohibit the propagation of buoyancy-restored gravity waves, which compose the baroclinic component of the atmospheric tidal response \citep[e.g.,][]{gerkema2008introduction}. This leaves the atmosphere with the barotropic component of the tidal flow, a feature consistent with tidal dynamics under the shallow water approximation (\ref{Appendix_non_dimensionization}).

Hereafter, we focus on the thermotidal heating as the only tidal perturber, and we ignore the much weaker gravitational potential $\tilde{U}$. It follows, in the neutral stratification limit, that $\gamma=0$ (Table \ref{tab:control_param}), and the vertical wavenumber (Eq. \ref{wavenumber2}) reduces to\footnote{It is noteworthy that the wavenumber in the neutral stratification limit is not longer dependent on the horizontal structure. }
\begin{equation}\label{wavenumber3}
    \hat{k}^2 = \left[\frac{1+\kappa +i\alpha}{2(\alpha-i)}\right]^2.
\end{equation}
It also follows that the background profiles of the scalar variables read as \citep{auclair2017atmospheric}:
\begin{align}\nonumber
    p_0(x) &= p_0(0)e^{-x}, \hspace{.5cm}
    \rho_0(x) = \frac{p_0(0)}{gH(0)}e^{(\kappa-1)x},\hspace{.3cm}\\\label{profiles}
    T_0(x) &= \frac{gH(0)}{\mathcal{R}_{\rm s}}e^{-\kappa x}.
\end{align}
We thus obtain for the heating profile (using Eqs. \ref{thermal_profile}, \ref{Jx_integral}, and \ref{delta_f_tot})
\begin{equation}
    J_{\rm s}=\delta F_{22} \frac{\alpha_{\rm A}g(b_{\rm J}+1)}{p_0(0)}\!\!\left[ 1- \mu_{\rm gr}\zeta\frac{1+si}{1+(1+si)\zeta}  \right].
\end{equation}
As such, the wave equation \eqref{wave_equation} is rewritten as
\begin{equation}\label{wave_equation2}
    \frac{d^2\Psi_n}{dx^2}+\hat{k}^2\Psi_n = \mathcal{A}_ne^{-\mathcal{B}x},
\end{equation}
where the complex functions $\mathcal{A}_n$ and $\mathcal{B}$ are defined as:
\begin{align}
&\mathcal{A}_n=\frac{\kappa\Lambda_n}{R_{\rm p}^2\sigma^3}\frac{\alpha^2+i\alpha}{\alpha^2+1}J_{\rm s},\\
&\mathcal{B}=\frac{1}{2(\alpha^2+1)}\left[ (2b_{\rm J}+1)(\alpha^2+1)-\kappa +i\kappa\alpha\right].
\end{align}
The wave equation \eqref{wave_equation2} admits the general solution
\begin{equation}\label{gen_solution}
    \Psi_n(x) = c_1e^{i\hat{k}x} + c_2e^{-i\hat{k}x} + \frac{\mathcal{A}_n}{\mathcal{B}^2+\hat{k}^2}e^{-\mathcal{B}x}.
\end{equation}
We consider the following two boundary conditions:
\begin{itemize}
    \item First, the energy of tidal flows, $\mathcal{W}$, should be bounded as $x\!\rightarrow\!\infty$. In \ref{appendix_energy_flow}, we derive  the expression of the tidal energy following \cite{Wilkes49}, and it scales as $\mathcal{W}\propto|\Psi|^2|\Phi|^2$. Accordingly, the non-divergence of the flow condition is satisfied if one sets $c_2=0$ and takes the proper sign of the wavenumber (Eq. \ref{wavenumber3}), namely:
    \begin{equation}
        \hat{k} = \frac{1}{2(\alpha^2+1)}\left[\kappa\alpha + i(1+\kappa+\alpha^2)\right].
    \end{equation}
    \item The second condition is the natural wall condition imposed by the ground interface, which enforces $\tilde{V}_{r; n}(x=0)=0$. We derive the expression of the profile of the vertical velocity in \ref{polarizations_appendix}, and by virtue of Eq.\eqref{vertical_velocity}, this condition allows us to write $c_1$ in the form:
    \begin{equation}\label{Def_A}
        c_1(x) = \frac{\mathcal{A}_n}{\mathcal{B}+{\hat{k}^2}}\times\frac{\mathcal{B} -\frac{1}{2}\left(1+\frac{i\kappa}{\alpha-i}\right)-\beta\Lambda_n+1}{i\hat{k} + \frac{1}{2}\left(1+\frac{i\kappa}{\alpha-i}\right) + \beta\Lambda_n -1}.
    \end{equation}
\end{itemize}
Under these boundary conditions, we are now fully geared to analytically compute the solution of the wave equation, $\Psi_n(x)$ (or equivalently $\tilde{G}_n(x))$, but we are specifically interested in retrieving a closed form solution of the quadrupolar tidal torque. The latter takes the general form\footnote{We note that this form corresponds to the quadrupolar component of the torque about the spin axis, and it is only valid assuming a thin atmospheric layer under the hydrostatic approximation. In the case of a thick atmosphere, one should integrate the mass redistribution over the radial direction. }  (\ref{Appendix_Torque}):
\begin{equation}\label{Equation_Torque}
    \mathcal{T} = \sqrt{\frac{6\pi}{5}}\frac{M_{\star}}{M_{\rm p}}\frac{R_{\rm p}^6}{a_{\rm p}^3}\Im\left\{\delta p_{\rm s}\right\}.
\end{equation}
Here $M_{\star}$ and $M_{\rm p}$ designate the stellar and planetary masses respectively, and $\Im$ refers to the imaginary part of a complex number, the latter in this case being the quadrupolar pressure anomaly at the surface $\delta p_{\rm s}=\delta p_{2}^{2,\nu}(x=0)$. \RVA{We further note that while this torque is computed for the atmosphere, it does act on the whole planet since the atmosphere is a thin layer that features no differential rotation with respect to the rest of the planet.}

Taking the solution $\Psi(0)$ of Eq. \eqref{gen_solution} (with $c_2=0$ and $c_1$ defined in Eq. \ref{Def_A}), we retrieve $\delta p_{\rm s}$ from Eq. \eqref{pressure_anomaly}. After straightforward, but rather tedious manipulations, we extract the imaginary part of the pressure anomaly and write it in the simplified form:
\begin{align}\label{imag_pressure_anomaly}\nonumber
    \Im\{\delta p_{\rm s}\} =& \alpha_{\rm A}\delta F_{22}\frac{\kappa g \Lambda_2}{R_{\rm p}^2 \sigma^3}\frac{(\mathcal{X}\alpha+\mathcal{Y})\alpha}{\left(1+2\zeta + 2\zeta^2\right)}\\
    & \times {\underbrace{\left[(\kappa-\beta\Lambda_2+1)^2+\alpha^2(\beta\Lambda_2-1)^2\right]}_\text{position of the Lamb resonance}}^{-1},
\end{align}
where we have defined the complex functions $\mathcal{X}$ and $\mathcal{Y}$ as 
\begin{align}\nonumber
    \mathcal{X}&=   (\beta\Lambda_2-1)\left[2\zeta^2(1-\mu_{\rm gr})+\zeta(2-\mu_{\rm gr})+1\right],\\ \label{Chi_Upsilon}
    \mathcal{Y}&=-s\mu_{\rm gr}\zeta(\kappa-\beta\Lambda_2 +1).
\end{align}
\RVA{We note that we provide the  full complex transfer function of the surface pressure anomaly, along with further analysis on its functional form in \ref{App_functional_Form}.} Before embarking on any results, we pause here for a few remarks on the provided closed form solution of the torque. 
\begin{figure}
\centering
\includegraphics[width=.45\textwidth]{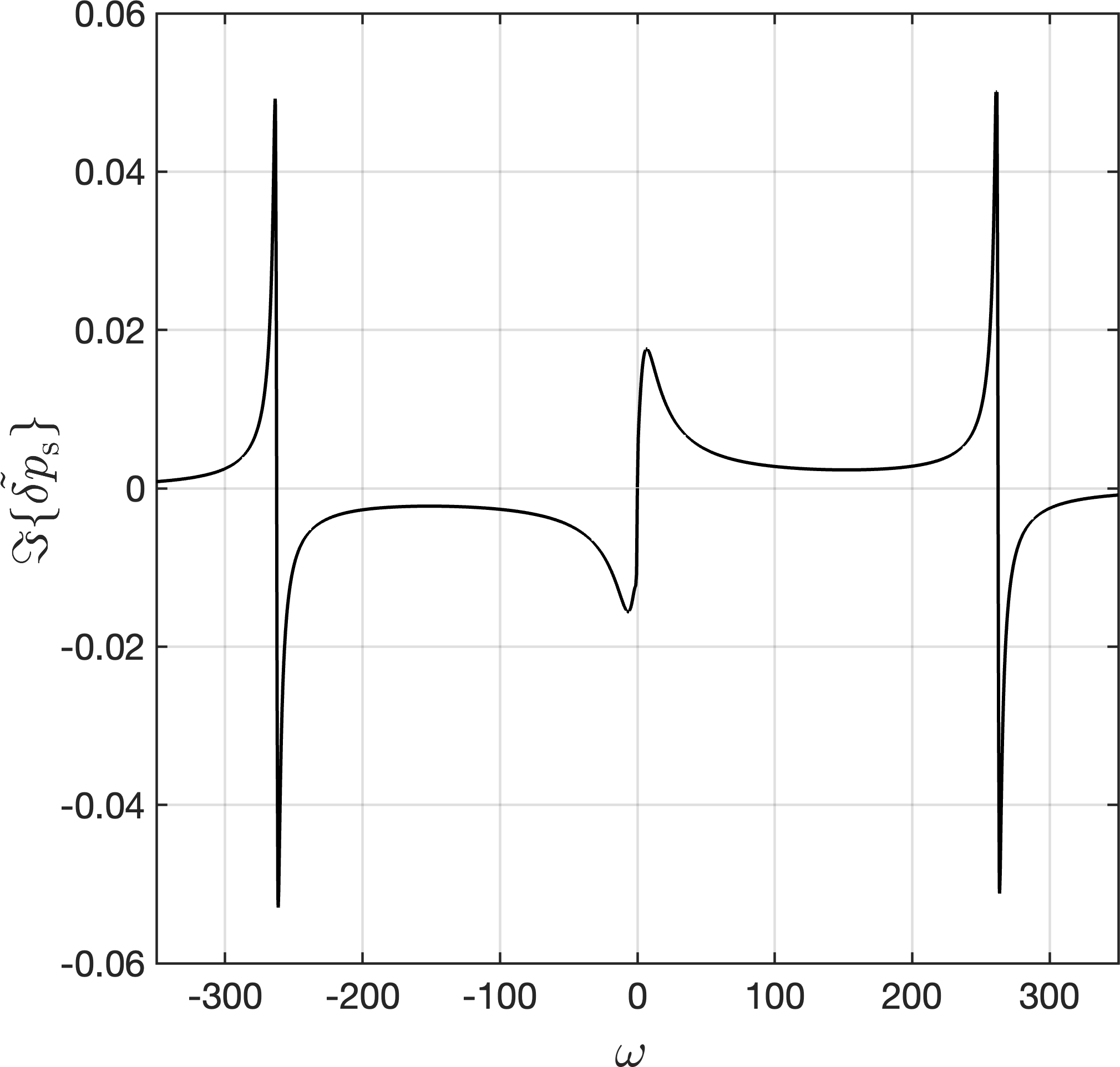}
    \caption{The spectrum of semi-diurnal atmospheric thermal  tides. Plotted is the imaginary part of the normalized pressure anomaly ($\tilde{\delta p}=\delta p/p_{\rm s}; $ Eq.\ref{imag_pressure_anomaly}) as a function of the normalized forcing frequency $\omega= (\Omega-n_\star)/n_\star=\sigma/2n_\star$, where $n_\star$ is the mean motion of the stellar perturber. The planetary-stellar parameters are those of the fiducial planetary system defined in Section \ref{Section_Response_NS}. }
    \label{Fig_Full_sym_spectrum}
\end{figure}

\begin{itemize}
    \item The parameter $\alpha_{\rm A}$, defined earlier (Eq.\ref{Jx_integral}) as the fraction of radiation actually absorbed by the atmosphere, can  evidently be correlated with the typical transmission function of the atmosphere and therefore its optical depth. Presuming that thermotidal heating on Earth is driven by ozone and water vapor, $\alpha_{\rm A}$ can then characterize atmospheric opacity parameter in the visible. Explicitly showing this dependence now takes us too far afield, though we compute and infer estimates of $\alpha_{\rm A}$ in Section \ref{Section_Resonance_Amplitude} and \ref{App_Alpha_A}.   
    \item The quadrupolar component of the equilibrium stellar flux, entering through a fraction of $F_\star$ (\ref{quadrupolar_forcing_appendix}), is directly proportional to the stellar luminosity $L_\star$. Standard models suggest that the Sun's luminosity was around 80\% of its present value ${\sim}3$~Ga \citep{gough1981solar}. Such luminosity evolution of Sun-like stars can be directly accommodated in the model if one were to study the evolution of the tidal torque with time.
    \item As we mentioned earlier, upon separating the horizontal and vertical structure of tidal dynamics, the only remaining coupling factor between the two structures is the eigenvalue of horizontal flows, $\Lambda_n$, in our case reducing to the dominant fundamental mode $\Lambda_2$. Noting that we have dropped the superscripts, we remind the reader that for the semidiurnal ($m=2)$ response, $\Lambda_2=\Lambda^{2,\nu}_{2}$, thus $\Lambda$ is frequency-dependent in the general case. The Earth however, over its lifetime, lives in the asymptotic regime of $\nu\approx1$ since $2\Omega\gg n_\star$, thus it is safe to assume that $\Lambda_2$ is invariant over the geological history\RVA{ with a value of 11.159 that we compute using the spectral method of \cite{wang2016computation}}. 
    \item Of significance to us in the Precambrian rotational equilibrium hypothesis is the tidal frequency, and consequently the LOD, at which the Lamb resonance occurs. It is evident from the closed form solution (\ref{imag_pressure_anomaly}) that the position of the resonance is controlled by the highlighted term. Had it not been for the introduced radiative losses, entering here through $\alpha$, this term would have encountered a singularity at the spectral position of the resonance, i.e. for $\beta\Lambda_2=1$. Here, however, the amplitude of the tidal peak is finite, and its position is a function of the planetary radius, gravitational acceleration, average surface temperature, eigenvalue of the fundamental Hough mode of horizontal flows, and the Newtonian cooling frequency. We detail further on this dependence in Section \ref{Section_LODres_position}.
\end{itemize}

In Fig. \ref{Fig_Full_sym_spectrum}, we plot the spectrum of the tidal response for a fiducial system in terms of the normalized surface pressure anomaly over a wide range of tidal frequencies covering the low and high frequency regimes. The system describes a Venus-like dry planet ($M_{\rm p}=0.815 M_{\oplus}, \,R_{\rm p}=0.95 R_{\oplus},\,a_{\rm p}=0.73\,{\rm au},\,g=8.87\,{\rm m\,s^{-2}})$, with a 10 bar atmosphere and a scale height at the surface $H_0=10$ km, thermally forced by a solar-like star ($M_\star=1M_\odot,\,L_\star=1L_{\odot})$. We further ignore here the thermal inertia in the ground and the atmosphere by taking $\sigma_{\rm bl}\rightarrow\infty$, or $\zeta\rightarrow0$, thus assuming an synchronous response of the ground with the thermal excitation.

\begin{figure*}[h]
\centering
\includegraphics[width=1\textwidth]{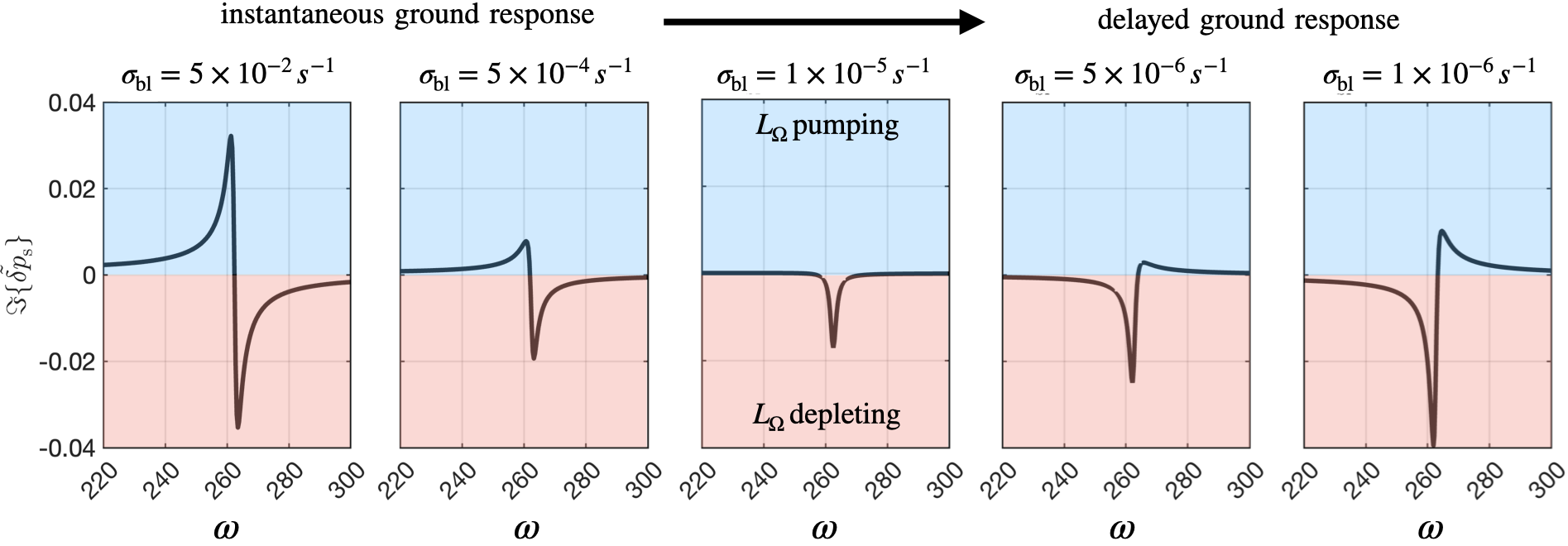}
    \caption{The (a-)symmetry of the Lamb resonance. Similar to Figure \ref{Fig_Full_sym_spectrum}, plotted is the imaginary part of the normalized pressure anomaly (Eq.\ref{imag_pressure_anomaly}), associated with the semidiurnal tide, as a function of the normalized forcing frequency $\omega= (\Omega-n_\star)/n_\star=\sigma/2n_\star$, for the same planetary-stellar parameters. We focus here on the high frequency regime around the Lamb resonance. Different panels correspond to different values of $\sigma_{\rm bl}$, or different thermal inertias in the ground and the atmosphere. Allowing for thermal inertia results in a delayed ground response, of which the signature is clear in inducing an asymmetry in the spectral behavior around the resonance.  }
    \label{Lamb_asymmetry}
\end{figure*}

Key tidal response features are recovered in this spectrum: First, we obtain a tidal peak near synchronization $(\omega=0)$ that generates a positive torque for $\sigma>0$ and a negative torque for $\sigma<0$, driving the planet in both cases away from its destined spin-orbit synchronization due to the effect of solid tides \citep[e.g.,][]{gold1969atmospheric,correia2001four,leconte2015asynchronous}. The peak has often been modelled by a Maxwellian functional form, though this form does not always capture GCM-generated spectra when varying the planetary setup \citep[e.g.,][]{auclair2019generic}. Second, we recover the Lamb resonance in the high frequency regime. The resonance is characterized here by two symmetric peaks of opposite signs. Thus upon passage through the resonance, the thermotidal torque shifts from being a rotational \RVR{pump to being a rotational brake}{brake to being a rotational pump}. In this work, we are more interested in the high frequency regime, thus we delegate further discussion and analysis on the low frequency tidal response to a forthcoming work, and we focus next on the Lamb resonance.

\subsection{The longwave heating limit: Breaking the symmetry of the Lamb resonance}\label{Section_Breaking_symmetry}
We now allow for variations of the characteristic time scale associated with the boundary layer diffusive processes, $\tau_{\rm bl}$ (Eq.\ref{eq_tau_bl}), or equivalently $\sigma_{\rm bl}$. Variations in $\sigma_{\rm bl}$ are physically driven by variations in the thermal conductive capacities of the ground and the atmosphere, and are significant when infrared ground emission and boundary layer turbulent processes contribute significantly to the thermotidal heating.

\begin{figure*}[t]
\centering
\includegraphics[width=1\textwidth]{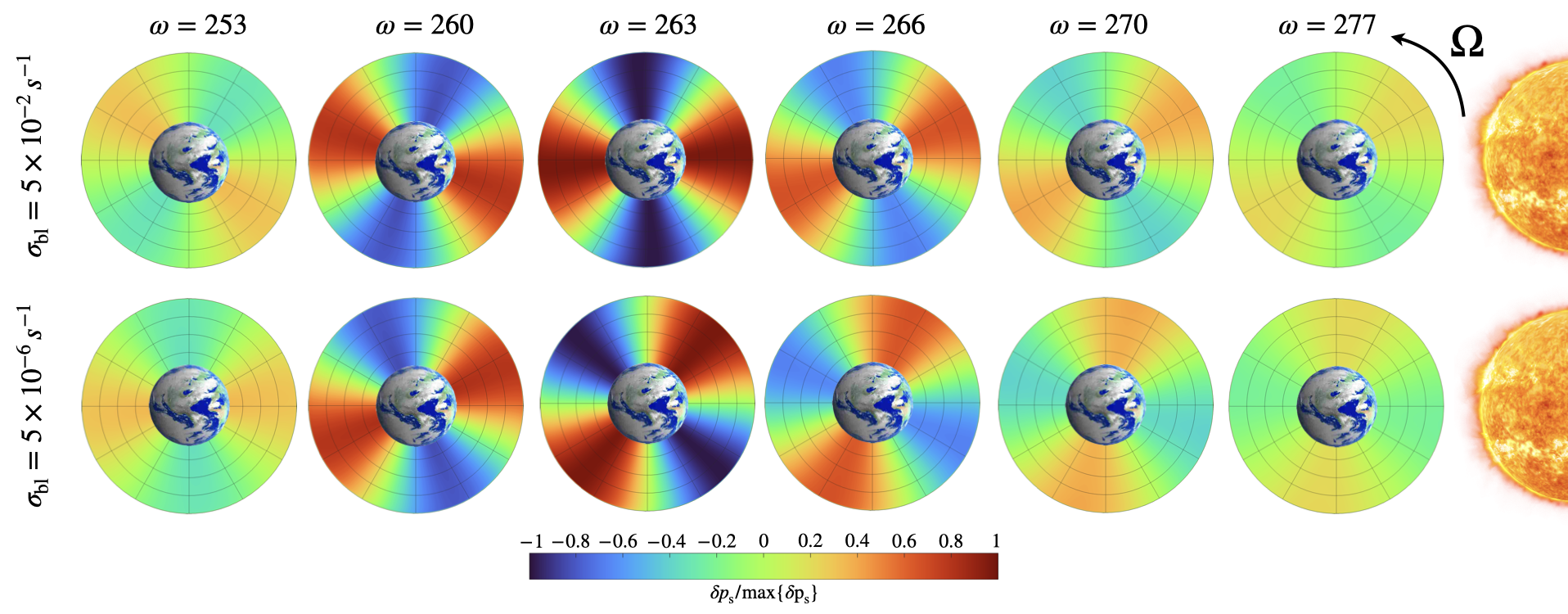}
    \caption{The thermally induced tidal bulge revealed. Shown are polar snapshots of the radial and longitudinal variations of the tidal pressure anomaly $\delta p(x)$ in the equatorial plane. The snapshots are shown from a top view, and the troposphere is puffed in size by virtue of the used mass-based vertical coordinate $\varsigma=p/p_{\rm s}$. The longitudinal axes are shown in increments of 30$^\circ$ with $0^\circ$ at the substellar point, while the radial axes are in increments of 0.25. \RVA{The profile of the pressure perturbation is also normalized by the exponentially decaying pressure background profile.} Snapshots are taken at different spectral positions that cover the passage through the Lamb resonance, \RVA{which specifically occurs at $\omega=262.6$}. In the top row, the response describes the limit of a planet with a synchronous atmospheric absorption, mimicking the Earth's direct absorption by ozone and water vapor, and it shows the continuous movement of the bulge, function of $\omega$, from lagging to leading the substellar point. In contrast, in the bottom row, and for the prescribed value of $\sigma_{\rm bl}$, the delayed response of the ground forces the bulge to always lag the substellar point, thus acting to decelerate the planetary rotation. }
    \label{Equatorial_maps_lag}
\end{figure*}

In such a case, the value of $\sigma_{\rm bl}$  plays a significant role in the tidal response of the planet. Namely, the ratio $\sigma/\sigma_{\rm bl}$ determines the angular delay of the ground temperature variations. For our study of the global tidal response, this frequency ratio determines whether the ground response is synchronous with the thermal excitation (when $\sigma\ll\sigma_{\rm bl})$, meaning thermal inertias vanish, the ground and the surface layer do not store energy, and the ground response is instantaneous; or if due to the combination of thermal inertias, the energy reservoir of the ground is huge, and the ground response lags the excitation, imposing another angular shift on the generated tidal bulge (when $\sigma\gtrsim\sigma_{\rm bl})$. We now reap from the analytical model the  signature of $\sigma_{\rm bl}$ in order to explain the Lamb resonance asymmetry -- as opposed to its symmetry in Figure \ref{Fig_Full_sym_spectrum} -- observed in GCM simulations of an atmosphere forced by a longwave flux \citep{auclair2019generic}.

In Figure \ref{Lamb_asymmetry}, we plot the tidal spectrum around the Lamb resonance, in terms of the normalized pressure anomaly at the surface, for different values of $\sigma_{\rm bl}$. For $\sigma_{\rm bl}=5\times10^{-2}$\,$s^{-1}$, the almost instantaneous response of the ground  leaves us with two pressure peaks that are symmetric around the resonant frequency. Decreasing $\sigma_{\rm bl}$ and allowing for a delayed ground response, the two pressure peaks of the resonance are attenuated in amplitude, but not with the same magnitude; namely, the amplitude damping is stronger against the positive pressure peak. Decreasing $\sigma_{\rm bl}$ to $10^{-5}\,s^{-1}$ in the panel in the middle, the positive pressure peak completely diminishes, leaving only the negative counterpart. Decreasing $\sigma_{\rm bl}$ further, both peaks are amplified, thus the positive peak emerges again. However, the spectral position of the peaks is now opposite to what it was in the limit of an instantaneous ground response. 

Given the direct proportionality between the tidal torque and the surface pressure anomaly (Eq.\ref{Equation_Torque}), the effect of thermal inertia thus contributes to the rotational dynamics when encountering the Lamb resonance. If a planet is decelerating and is losing rotational angular momentum, $L_{\Omega}$, due to solid or oceanic gravitational tides, $\omega$ decreases, and the planet encounters the resonance from the right. In the first panel of Figure \ref{Lamb_asymmetry}, the thermotidal torque in this regime is also negative, thus it complements the effect of gravitational tides. When the resonance is encountered, the thermotidal torque shifts its sign to counteract the effect of gravitational tides, with an amplified effect in the vicinity of the resonance. However, with the introduction of thermal inertia into the linear theory of tides, the $L_\Omega$-pumping part of the atmospheric torque is attenuated, and for some values of $\sigma_{\rm bl}$, it completely disappears. This modification of the analytical theory allows us to explain the asymmetry of the Lamb resonance depicted in the 3D GCM simulations of \cite{auclair2019generic}. In \ref{appendix_fitting_PAD19}, we show that we are able to recover from our model the essential features of the tidal spectrum computed in the mentioned simulations. 

To understand the signature of the surface response further, in Figure \ref{Equatorial_maps_lag}, we generate snapshots of the tidal pressure variation in the equatorial plane, seen from a top view. The snapshots thus show the thermally induced atmospheric mass redistribution and the resulting tidal bulge, if any. To generate these plots, we compute the vertical profile of the pressure anomaly from Eq. \eqref{pressure_anomaly}, and augment it with the latitudinal and longitudinal dependencies from Eqs. (\ref{Fourier_expansion}-\ref{Hough_expansion}). As the massive troposphere dominates the tidal mass redistribution, we use the mass-based vertical coordinate $\mathcal{\varsigma}=p/p_{\rm s}$ (i.e. $x=-\ln\varsigma$, and $\varsigma$ ranges between $1$ at the surface and $0$ in the uppermost layer).

In  Figure \ref{Equatorial_maps_lag}, we show the tidal bulge as the planet passes through the Lamb resonance, for two values of $\sigma_{\rm bl}$ that correspond to the limits of synchronous atmospheric absorption (top row), and a delayed thermal response in the ground (bottom row). First, the accumulation of mass and its culmination on a tidal bulge is indicated by the color red, with varying intensity depicting varying pressure amplitudes. In the case of synchronous atmospheric absorption, for $\omega=253$, i.e. before encountering the resonance, the bulge leads the substellar point and acts to accelerate the planet's rotation. Increasing $\omega$ and encountering the resonance, the bulge reorients smoothly towards lagging the substellar point thus decelerating the planet's rotation. This behavior is consistent with the established response spectrum in the first panel of Figure \ref{Lamb_asymmetry}, and is relevant to the Earth's case, assuming that thermotidal heating is predominantly driven by direct synchronous absorption. In the bottom row, the delayed response of the ground imposes another shift on the bulge: for the prescribed value of $\sigma_{\rm bl}$, the passage through the resonance only amplifies the response, but the bulge barely leads the tidal vector, leaving us with a tidal torque that mainly complements the gravitational counterpart, as seen in the fourth panel of Figure \ref{Lamb_asymmetry}.

From what preceded, the reader can find it quite natural that the effect of thermal inertias in the ground and the boundary layer should be accounted for when studying planetary rotational dynamics using the linear theory, especially under  longwave forcing.  The results also make it tempting to revisit these effects in the case of the dominant shortwave forcing on Earth, as they have been often ignored from the theory \citep[e.g.,][]{chapman1969atmospheric} on the basis of the small-amplitude non-migrating tidal components they produce \citep[e.g.,][]{schindelegger2014surface}.

\section{A fixed Precambrian LOD for the Earth?}\label{Section_fixed_LOD?}
So where does all this leave us with the Precambrian rotational equilibrium hypothesis? The occurrence of this scenario straddles several factors,  the most significant of which is that the Lamb resonance amplifies the thermotidal response when the opposing gravitational tide is attenuated. Consequently, to investigate the scenario, the two essential quantities that need to be well constrained are the amplitude of the thermotidal torque when the resonance is encountered, and the geological epoch of its occurrence. Having provided a closed form solution for the tidal torque, it is straightforward for us to investigate these elements. 

\subsection{Was the resonance resonant enough? A parametric study}\label{Section_Resonance_Amplitude}

Constraints on the amplitude of the gravitational tide during the Precambrian are model-dependent. The study in \cite{zahnle1987constant}, and later in \cite{bartlett2016analysis}, relied on rotational deceleration estimates fitted to match the distribution of geological proxies available at the time \citep[e.g.,][]{lambeck2005earth}. Specifically, the estimate of the  Precambrian gravitational torque relied on the tidal rhythmite record preserved in the Weeli-Wolli Banded Iron formation \citep{walker1986lunar}. The record is fraught with multiple interpretations featuring different inferred values for the LOD \citep{williams1990tidal,williams2000geological}, altogether different from a recent cyclostratigraphic inference that roughly has the same age \citep[][see Figure \ref{LOD_evo_models} for the geological data points ${\sim}2.45$ Ga]{lantink2022milankovitch}. \RVA{Nevertheless, the claim of an attenuated Precambrian torque still holds, as  the larger interval of the Precambrian is associated with a ``dormant'' gravitational torque phase, lacking any significant amplification in the oceanic tidal response, in contrast with the present state where the oceanic response lives in the vicinity of a spectral resonance \citep[e.g.,][]{farhat2022resonant}}.

\begin{figure}[t]
\centering
\includegraphics[width=.45\textwidth]{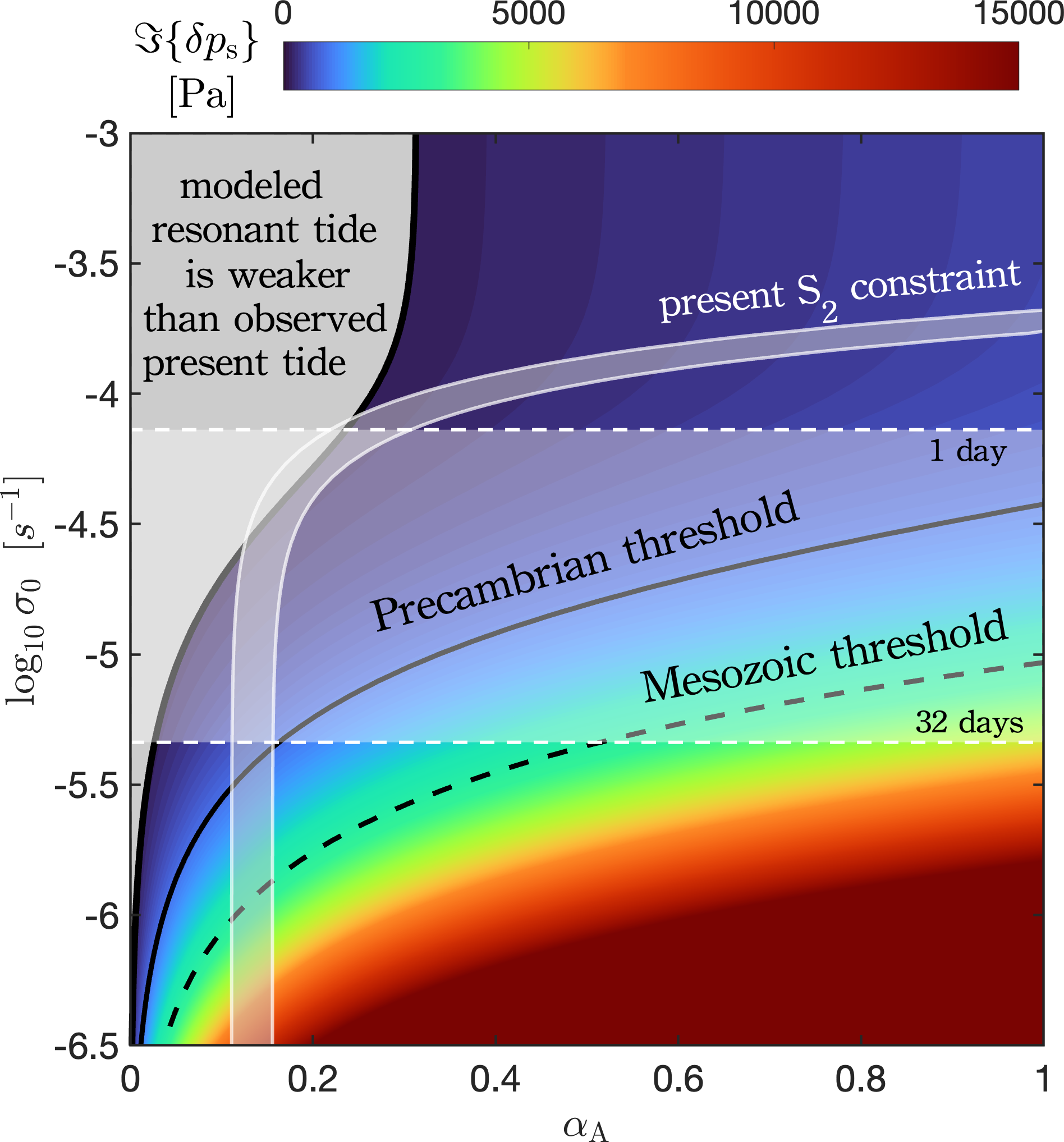}
    \caption{A parametric study of the tidal response. Plotted is a contoured surface of the amplitude of the imaginary part of the positive semidiurnal pressure anomaly at the Lamb resonance, over a grid of values of our free parameters $\sigma_0$ and $\alpha_{\rm A}$. The solid black isoline marks the level curve of $\Im\{\delta p_{\rm s}\}=880 $ Pa, and defines from below a region in ($\alpha_{\rm A},\sigma_0)$-space where the thermotidal response is sufficient to cancel the gravitational counterpart in the Precambrian. Analogously, the dashed isoline defines the threshold ($\Im\{\delta p_{\rm s}\}=2275 $ Pa) needed in the early Mesozoic, 250 Ma.  The horizontal shaded area corresponds to typical values of the radiative cooling rate as described in the main text. The other shaded area defines the region of parameter space that yields the presently observed semi-diurnal tidal bulge. The gray area on the left covers the parametric region where the resonance features a lower pressure amplitude than the present.}
    \label{pressure_map_Earth_alphaA}
\end{figure}

That said, we explore the atmospheric parameter space of our analytical model to check the potential outcomes of the torques' competition. Given that the dominant thermotidal forcing on Earth is the direct absorption of the incident flux, we consider the synchronous limit of $\zeta\rightarrow0$, whereby the Lamb resonance is symmetric (first panel of Figure \ref{Lamb_asymmetry}; top row of Figure \ref{Equatorial_maps_lag}). In Figure~\ref{pressure_map_Earth_alphaA}, on a grid of values of our free parameters $\sigma_0$ and $\alpha_{\rm A}$, we contour the surface of the maximum value of the imaginary part of the positive pressure anomaly that is attained when the Lamb resonance is encountered. The two parameters have a similar signature on the tidal response.  Moving vertically upwards and increasing the rate of Newtonian cooling typically attenuates the amplitude of the peak. For very high cooling rates corresponding to $\sigma_0\gtrsim10^{-4}\,s^{-1}$, we severely suppress the amplified pressure response around the resonance. Conversely, for values of $\sigma_0\lesssim10^{-6.5}\,s^{-1}$, we approach the adiabatic limit of the tidal model where the Lamb resonance becomes a singularity. A similar signature is associated with increasing the opacity parameter of the atmosphere. 

On the contour surface, we highlight with the solid black isoline the pressure anomaly value required to generate a thermotidal torque of equal magnitude to the Precambrian gravitational tidal torque. The latter (${\sim}1.13\times 10^{16}$ N m) is roughly a quarter of the present gravitational torque (${\sim}4.51\times 10^{16}$~N~m) \citep[e.g.,][]{zahnle1987constant,farhat2022resonant}, thus requiring, via Eq.\eqref{Equation_Torque}, $\Im\{\delta p_{\rm s}\}$ on the order of 880 Pa\footnote{This value does not correspond to the maximum amplitude of the surface pressure oscillation at the equator, but to the coefficient of its quadrupolar spherical harmonic component, which via \ref{quadrupolar_forcing_appendix} and \ref{Appendix_Torque} is normalized by a factor of $\sqrt{16\pi/15}$. 
}.  This isoline bounds from below a cornered region of the parameter space where the thermal tide is sufficiently amplified upon the resonance encounter.  It is noteworthy that this Precambrian value of the torque is the minimum throughout the Earth's history. We mark by the dashed isoline, for comparison, the threshold needed if the Lamb resonance is encountered in the Mesozoic ($\Im\{\delta p_{\rm s}\}=2275$ Pa). The solid gray region on the left side of the parameter space is bounded by the isoline corresponding to the present value $\Im\{\delta p_{\rm s}\}=224$~Pa \citep{schindelegger2014surface}. Thus it defines to the left an area where the present thermal tide is stronger than it would be around the resonance.

We take this parametric study one step further to study whether typical values of the parameters $\sigma_0$ and $\alpha_{\rm A}$ can place the Earth's atmosphere in the identified regions. Stringent constraints on $\sigma_0$ are hard to obtain for the Earth since $\sigma_0$ is an effective parameter that in reality is dependent on altitude. Furthermore, in the linear theory of tides, we are forced to ignore the layer-to-layer radiative transfer and assume a gray body atmospheric radiation directly into space. However, radiative transfer can be consistently accommodated in numerical GCMs using the method of correlated k-distributions \citep[e.g.,][]{lacis1991description} as performed in \cite{leconte2015asynchronous} and in \cite{auclair2019generic}, both studies using the LMD GCM \citep{hourdin2006lmdz4}. In fact, \cite{leconte2015asynchronous} fitted their numerically obtained atmospheric torques to  effective values of $\sigma_0$ for various atmospheric parameters \citep[see Table 1 of][]{leconte2015asynchronous}. The closest of these settings  to the Earth yields a radiative cooling timescale  $\tau_{\rm rad}= 32$ days. In contrast, \cite{lindzen1968application} and later \cite{lindzen1972lamb} estimated the timescale to be on the order of 1 day. We presume that these estimates should encompass the possible effective values for the Earth's atmosphere, and we highlight with the horizontal shaded area the range of these values\footnote{We note that in \cite{leconte2015asynchronous}, the tidal frequency under study is that of the diurnal component, thus we multiply their $\omega_0$ value by 2; i.e. $\sigma_0=2\omega_0$.}.

Another constraint on the free parameters emerges from present  in situ barometric observations of the semidiurnal ($S_{2}$) tidal response. We use the analysis of compilations of measurements performed in \citet{haurwitz1973diurnal,dai1999diurnal,covey2014atmospheric} and \cite{schindelegger2014surface}, which constrain the amplitude of the semi-diurnal surface pressure oscillation to within $107-150$ Pa, occurring around 0945~LT. The narrow shaded area defines the region of parameter space that can explain these observables using the present semi-diurnal frequency, placing the opacity parameter in the region $\alpha_{\rm A}{\sim}14\%$. In \ref{App_Alpha_A}, we compute estimates of the present value of $\alpha_{\rm A}$ by studying distributions of heating rates that are obtained either by direct measurements of the Earth's atmosphere \citep{chapman1969atmospheric}, or using GCM simulations \citep{vichare2013diurnal}. Our analysis of the data suggests that the efficiency parameter is around $\alpha_{\rm A} {\sim}17{-}18\%$, which is consistent with the $S_2$ constraint we obtain. \RVA{Finally, it is also noteworthy how the plotted $S_2$ constraint is insensitive to variations in $\sigma_0$ over a wide interval, which prohibits the determination of the present value of $\sigma_0$ using this constraint.}

Evidently, the overlap of the parametric constraints lives outside the  region where the thermotidal response is sufficient for the rotational equilibrium condition. \RVA{The present thermotidal torque ($2.89\times10^{15}$ N m) needs to be amplified by a factor of 3.9 to reach the absolute minimum of the opposing gravitational torque\footnote{subject to the uncertainty of the present measurement of the semi-diurnal surface pressure oscillation discussed earlier} in the Precambrian, and by a factor of 12.3 to reach the Mesozoic value. Our parametric exploration precludes these levels of amplification. It is important to also note that larger amplification factors would be required if one were focused on the modulus of the pressure oscillation, rather than its imaginary part. This derives from Figure~\ref{pressure_components} where we show that the amplification in the imaginary part is almost half that of the modulus of the surface pressure oscillation.}

One can argue, however, that the used constraints derive from present measurements, and the likelihood of the scenario still hinges on possible atmospheric variations as we go backwards in time. \RVA{Nonetheless, the radiative cooling timescale exhibits a strong dependence on the equilibrium temperature of the atmosphere  \citep[$\sigma_0\propto T_0^3$;][Eq. 17]{auclair2017atmospheric}. As such, a warmer planet in the past would yield a shorter cooling timescale, and consequently, more efficient damping of the resonant amplitude (see \ref{Appendix_compositional}). On the other hand, atmospheric compositional variations can change the opacity parameter of the atmosphere in the visible and the infrared. An increase of the opacity in the visible to $\alpha_{\rm A}=24\%$ can indeed place the response beyond the Precambrian threshold for some values of $\sigma_0$. An increase to four times the present value of $\alpha_{\rm A}$ is required to cross the threshold in the Mesozoic. These increases, however, can be precluded, based on the fact that the Archean lacked a stratospheric ozone layer \citep[e.g.,][]{catling2020archean}. In contrast, an increase in atmospheric opacity in the infrared, which accompanies the abundance of Precambrian greenhouse gases, delivers the opposite effect by attenuating the resonant tidal response, as we elaborate in \ref{Appendix_compositional}. Furthermore, the latter increase would also trigger the contribution of asynchronous tidal heating, which further attenuates the amplitude of the positive peak as we show in Section \ref{Section_Breaking_symmetry}. Thus, with these analyses, it is unlikely that the resonance could have amplified the thermotidal response beyond the required threshold. This conclusion can be further regarded as conservative, since the employed linear model tends to overestimate the resonant amplification of the tidal response. This derives from the fact that, in the quasi-adiabatic regime, the model ignores the associated non-linearities of dissipative mechanisms. The remaining question is therefore: when did the Lamb resonance actually occur?}

\begin{figure}[t]
\centering
\includegraphics[width=.48\textwidth]{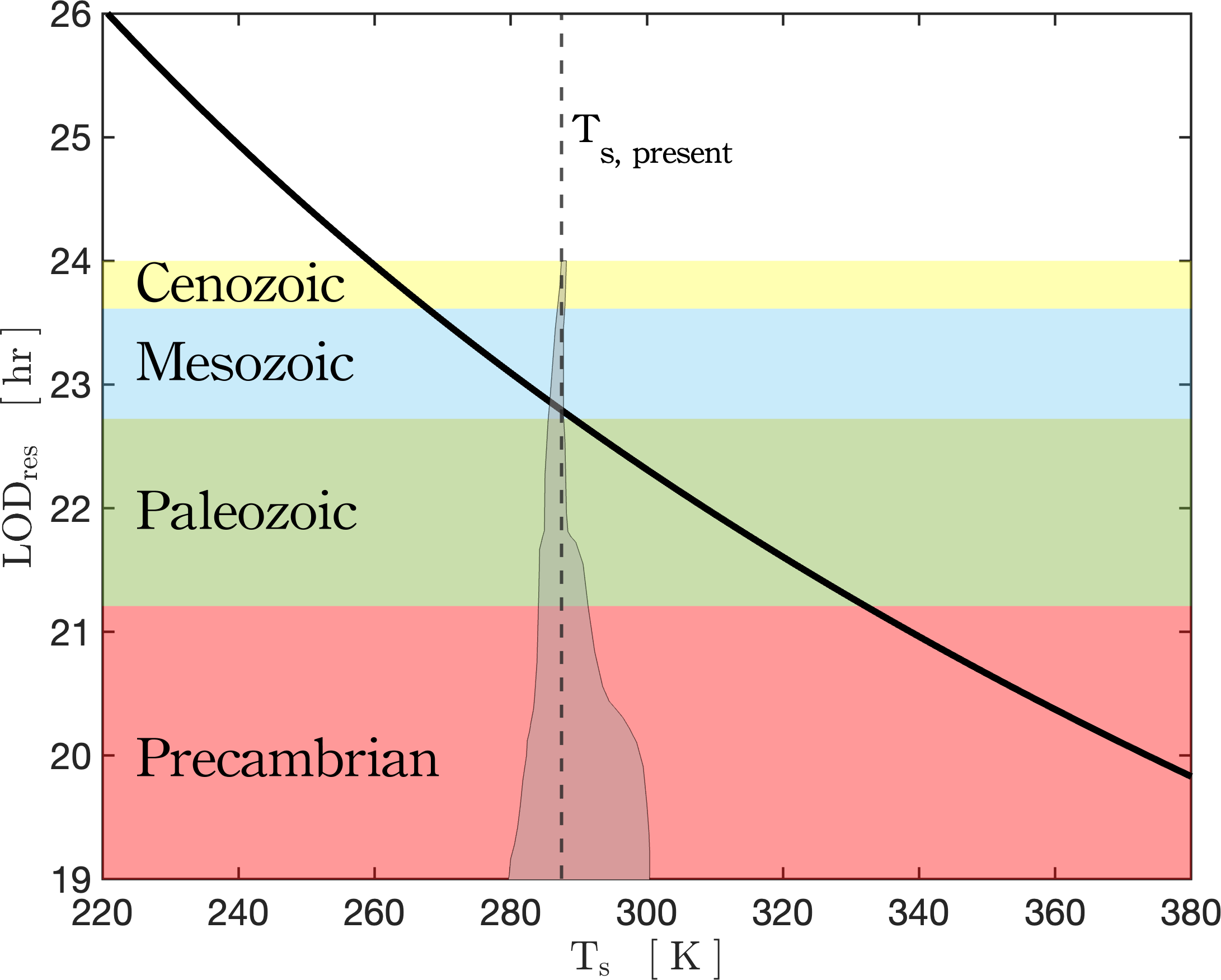}
    \caption{The dependence of the resonant rotational period on the mean surface temperature. By virtue of Eq.\eqref{eq_LOD_Ts}, the LOD at which the Lamb resonance occurs scales as the inverse square root of the mean surface temperature. The gray shaded area highlights 95\% confidence intervals for the past temperature evolution according to the carbon cycle model of \cite{krissansen2018constraining}. The identified geological eras correspond to the LOD evolution model of \cite{farhat2022resonant}. The overlap between the modelled temperature evolution and the black curve places the resonance occurrence in the early Mesozoic.}
    \label{LOD_Ts}
\end{figure}

\subsection{The spectral position of the Lamb resonance}\label{Section_LODres_position}

The spectral position of the Lamb resonance, or equivalently, the geological time of its occurrence, is identified in the analytical model via the denominator highlighted in Eq.\eqref{imag_pressure_anomaly}. The latter is a function of $\sigma$ and is dependent on the planetary radius, gravitational acceleration, eigenvalue of the fundamental Hough mode, the radiative frequency, and the equilibrium surface temperature at the surface $T_{\rm s}$, and is independent of $\sigma_{\rm bl}$. Thus for the Earth, the resonance position is merely dependent on the equilibrium temperature at the surface and the radiative cooling frequency.

In Figure \ref{LOD_Ts}, we plot the dependence of the spectral position of the resonance, in terms of LOD, on $T_{\rm s}$. The apparent single curve is actually a bundle of curves with different values of $\sigma_0$, but the effect of the latter is unnoticeable (if one varies $\sigma_0$ by two orders of magnitude, the resonant rotational period varies by few minutes). As such, the resonant frequency is predominantly controlled by $T_{\rm s}$, which allows us to take the adiabatic limit of Eq.\eqref{imag_pressure_anomaly}, and straightforwardly derive the tidal frequency that minimizes the denominator. In terms of the rotational period, the position of the resonance then reads:
\begin{equation}\label{eq_LOD_Ts}
    {\rm LOD_{res}}= \frac{4\pi R_{\rm p}}{\sqrt{\mathcal{R_{\rm s}}\Lambda_n T_{\rm s}}+2R_{\rm p}n_\star}.
\end{equation}
The resonant rotational period thus scales as the inverse square root of the surface equilibrium temperature. However, the evolution of the latter for the early Earth is widely debated. For instance, marine oxygen isotopes have been interpreted to indicate Archean ocean temperatures around $60-80^\circ$C \citep[e.g.,][]{knauth2005temperature,robert2006palaeotemperature}. This interpretation is in contrast with geochemical analysis using phosphates \citep[e.g.,][]{blake2010phosphate}, geological evidence of Archean glacial deposits \citep[e.g.,][]{de20163}, geological carbon cycle models \citep[e.g.,][]{sleep2001carbon,krissansen2018constraining}, numerical results of 3D GCMs \citep[e.g.,][]{charnay2017warm}, and the fact that solar luminosity was 10-25\% lower during the Precambrian \citep[e.g.,][]{charnay2020faint}, altogether predicting a temperate climate and moderate temperatures throughout the Earth's history.

We highlight with the gray shading on top of the curve  modelled mean surface temperature variations adopted from \cite{krissansen2018constraining}. As the latter temperature evolution is established in the time domain, we use the LOD evolution in \cite{farhat2022resonant} to map from time-dependence to LOD-dependence, and we further identify the corresponding geological eras of the LOD evolution with the color shadings.  Given the present day equilibrium surface temperature, the resonance occurs at LOD $=22.8\,{\rm hr}$. \RVA{This value is in  agreement with the $11.38\pm0.16\,{\rm hr}$  semi-diurnal period obtained by analyzing the spectrum of normal modes using pressure data on global scales \citep[see Table 1 of][first symmetric gravity mode of wavenumber $k=-2$]{sakazaki2020}. In \ref{App_isothermal_LOD}, we compute the resonant rotational period assuming an isothermal profile of the atmosphere, and we show that it is roughly one hour less than that in the neutrally stratified limit, placing it closer to $21.3\,{\rm hr}$ estimate of  \cite{zahnle1987constant} and 
\cite{bartlett2016analysis}}. \RVA{We emphasize here, however, that the resonant period does not exactly mark the period at which the thermotidal torque is maximum. The latter occurs at the peaks surrounding the resonance (see Figures \ref{Fig_Full_sym_spectrum} and \ref{pressure_components}), the difference between the two being dependent on the radiative cooling frequency.}

Taking the LOD evolution model of \cite{farhat2022resonant} at face value, the temperature variations predicted in \cite{krissansen2018constraining} locate the resonance encounter in the Triassic, and not in the Precambrian. In fact, for the resonance to be encountered in the Precambrian, even in the latest eras of it, the resonant period should move to less than ${\sim}21\,{\rm hr}$, but this requires an increase in the equilibrium temperature of  at least $55^\circ $C, which is inconsistent with the studies mentioned above. \RVA{Such an increase in temperature would also increase $\sigma_0$ by almost $19\%$ (as we discuss in the previous section, $\sigma_0\propto T_0^3$; see also \ref{Appendix_compositional}), reducing the radiative cooling timescale and prompting more efficient damping of the tidal amplitude at the resonance. Moreover, such an increase in temperature would most probably accompany increased greenhouse effects in the past, which in turn would increase the atmospheric absorption and thermotidal heating in the infrared. The latter would then place the Earth's atmosphere in the regime of asynchronous thermotidal heating studied in Section \ref{Section_Breaking_symmetry}, whereby the accelerative peak of the torque is further attenuated.}  

\section{Summary and Outlook}
We were drawn to the problem of atmospheric thermal tides by the hypothesized scenario of a constant length of day on Earth during the Precambrian. Our motivation in investigating the scenario lies in its significant implications on paleoclimatic evolution, and the evident mismatch between LOD geological proxies and the predicted LOD evolution if this rotational equilibrium is surmised. The scenario hinges on the occurrence of a Lamb resonance in the atmosphere whereby an amplified thermotidal torque would cancel the opposing torque generated by solid and oceanic gravitational tides. Naturally, the atmospheric tidal torque is that of two flavors: it can either pump or deplete the rotational angular momentum budget of the planet, depending on the orientation of the generated tidal bulge. 

With this rotational equilibrium scenario in mind, we have developed a novel analytical model that describes the tidal response of thermally forced atmospheres on rocky planets. The model derivation is based on the secure ground of the first principles of linear atmospheric dynamics, studied under classical approximations that are commonly drawn in earlier analytical works and in more recent numerical frameworks. The distinct feature that we imposed in this model is that of neutral atmospheric stratification, which presents a more realistic description of the Earth's troposphere than the isothermal profile imposed in earlier analytical studies. In this limit, we derive from the model a closed form solution of the tidal torque that can be efficiently used to study the evolution of planetary rotational dynamics. We accommodate into the model dissipative thermal radiation via linear Newtonian cooling, and turbulent and diffusive processes related to thermal inertia budgets in the boundary layer and the ground. As such, the model can be used to study a planetary thermotidal response when heated either by direct synchronous absorption of the incident stellar flux, or by a delayed infrared radiation from the ground.

We probed the spectral behavior of the tidal torque using this developed model in the two aforementioned limits. In the limit of longwave heating flux, the inherently delayed thermal response in the planetary boundary layer maneuvers the tidal bulge in such a way that, for typical values of thermal inertia in the ground and atmosphere, the  accelerating effect of the tidal torque at the Lamb resonance is attenuated, and possibly annihilated. In the case of the Earth, \RVA{ -- where we apply the opposite limit of shortwave thermotidal heating and ignore the attenuating effect of asynchronous forcing --} while the encounter of the resonance in the atmosphere is guaranteed, the epoch of its occurrence and the tidal amplitude it generates are uncertain. As such, we attempted a cautious incursion on constraining them and learned that:
\begin{itemize}
    \item Assuming that temperate climatic conditions have  prevailed over the Earth's history, the resonance is likely to have occurred in the early Mesozoic, and not in the Precambrian. The early Mesozoic, unlike the Precambrian, is characterized by an amplified decelerating luni-solar gravitational torque.  
    \item For judiciously constrained estimates of our atmospheric model parameters, the resonance does not amplify the accelerating thermotidal torque to a level comparable in magnitude to the gravitational counterpart.
\end{itemize}
These model predictions presume that thermotidal heating in the Earth has always been dominated by the shortwave. Compositional variations however, namely those associated with increased greenhouse contributions in the past would  amplify the asynchronous thermotidal forcing in the longwave. The latter in turn, as we show in this work, further attenuates the accelerating flavor of the resonant torque. Exploring this end is certainly worthy of future efforts, \RVA{but with the present indications at hand, we conclude that the occurrence of the rotational equilibrium is contingent upon a drastic increase in the Earth's surface temperature ($\geq55^\circ$C), a long enough radiative cooling timescale ($\geq 40$ days), an increase in the shortwave flux opacity of the atmosphere, and that infrared thermotidal heating remained negligible in the past. We cannot completely preclude these requirements when considered separately, especially given the uncertainty in reconstructing the Earth's temperature evolution in the Proterozoic. However, a warmer paleoclimate goes hand in hand with a shorter radiative cooling timescale, along with increased greenhouse gases that amplify the asynchronous thermotidal forcing. Both effects damp the accelerating flavor of the thermotidal torque. Put together, these indications suggest that the occurrence of the rotational equilibrium for the Earth is unlikely. To that end, future GCM simulations that properly model the Precambrian Earth to provide stringent constraints on our analytical predictions of the resonant amplification are certainly welcome.}

Ultimately though, \RVA{even if the locking into the resonance did not occur,} the effect of the thermotidal torque at the resonance remains a robust and significant feature, and it should be accommodated in future modelling attempts of the Earth's rotational evolution. Our model sets the table for efficiently studying such a complex interplay between several tidal players, both for the Earth and duly for its analogues. Interestingly, the question of the climatic response to the Lamb resonance, or similarly to oceanic tidal resonances, where abrupt and significant astronomical variations occur, largely remains an unexplored territory, perhaps requiring an armada of rigorous GCM simulations. This only leaves us with anticipated pleasure in weaving yet another thread in the rich tidal history of the Earth. 
Furthermore, we anticipate that the growing abundance of geological proxies, especially robust inferences associated with cyclostratigraphy, may help detect the whereabouts of these resonances and provide further constraints to our modeling efforts.

\section*{\textit{Acknowledgments}}
M.F. expresses his gratitude to Kevin Heng for his hospitality at the LMU Munich Observatory where part of this work was completed. This work has been supported by the French Agence Nationale de la Recherche (AstroMeso ANR-19-CE31-0002-01) and by the European Research Council (ERC) under the European Union’s Horizon 2020 research and innovation program (Advanced Grant AstroGeo-885250). This work was granted access to the HPC resources of MesoPSL financed by the Region Île-de-France and the project Equip@Meso (reference ANR-10-EQPX-29-01) of the  programme Investissements d’Avenir supervised by the Agence Nationale pour la Recherche. 


\appendix
\section{The dimensionless governing equations of atmospheric tidal dynamics}
\label{Appendix_non_dimensionization}
Using the same definition of atmospheric variables as in the main text, we develop here the dimensionless governing system of equations describing tidal dynamics, which shall be used to recover the wave equation (\ref{wave_equation}). We start with the classical primitive equations describing the atmospheric tidal response under the thin shell approximation \citep[e.g.,][]{siebert1961atmospheric,chapman1969atmospheric}, and we augment  the  heat transfer equation by the additional radiative term $J_{\rm rad}$ described in the main text. These equations read:
\begin{align}
 & \partial_{{t}}{V}_\theta -2\Omega\cos\theta{V}_{\varphi} = -\frac{1}{R_{\rm p}}\partial_{\theta}\left(\frac{\delta p}{\rho_0} - {U}\right)\,, \label{momentum_theta_App}\\
&  \partial_{{t}}{V}_\varphi \!\!+2\Omega\cos\theta{V}_{\theta} = \!-\frac{1}{R_{\rm p}\sin\theta}\partial_{\varphi}\!\left(\!\frac{\delta p}{\rho_0}\! - \!{U}\!\right)\!,\label{momentum_phi_App}\\
&0=-\partial_z\delta p-g\delta\rho+\rho_0\partial_zU\,,\label{momentum_r_App}\\
&\frac{D\rho}{Dt}=-\rho_0\chi\,,\label{mass_cons_App}\\
&\frac{1}{\Gamma_1}\frac{Dp}{Dt}- g H\frac{D\rho}{Dt}=\kappa\rho_0J-\rho_0\mathcal{R}_{\rm s}\sigma_0\delta T\,,\label{heat_eqn_App}\\
&\chi-\partial_z V_r = \nabla_{\rm h}\cdot \boldsymbol{V}\,,\\
&\frac{\delta p}{p_0} = \frac{\delta \rho}{\rho_0} + \frac{\delta T}{T_0}.\label{ideal_gas_law_App}
\end{align}
This system describes atmospheric momentum conservation (Eqs.~\ref{momentum_theta_App}~-~\ref{momentum_r_App}),  mass conservation (Eq.~\ref{mass_cons_App}), heat transfer with Newtonian cooling (Eq.~\ref{heat_eqn_App}), and the ideal gas law (Eq.~\ref{ideal_gas_law_App}). Furthermore, we have adopted  the traditional approximation \citep[e.g.,][]{unno1989nonradial}, where we ignore the Coriolis coupling term between the vertical and horizontal parts of the momentum equations, thus allowing for the analytical treatment of the system. 
For the vertical momentum equation (\ref{momentum_r_App}), the traditional approximation amounts to imposing the hydrostatic equilibrium for the tidal perturbation in addition to the background profiles, which are in turn averaged over the sphere. The hydrostatic approximation would allow us later to define the tidal torque as a function of the pressure anomaly at the surface. 

In this system, the material (read Lagrangian) time derivative $\frac{D}{Dt}$ of any variable $y$ is defined as
\begin{equation}
    \frac{Dy}{Dt} = \partial_t y + V_r \frac{dy_0}{dz},
\end{equation}
and we denote by $\chi$ the divergence of the velocity vector field $\boldsymbol{V}$, and by the notation $\nabla_{\rm h}\cdot \boldsymbol{V}$ its horizontal divergence such that
\begin{equation}
    \nabla_{\rm h}\cdot \boldsymbol{V} = \frac{1}{R_{\rm p}\sin\theta}\left[\partial_\theta(\sin\theta V_\theta)+\partial_\varphi V_\varphi\right].
\end{equation}
We also introduce the calculation variable 
\begin{equation}\label{Def_G_AppA}
    G = -\frac{1}{\Gamma_1 p_0} \frac{Dp}{Dt}.
\end{equation}
It is noteworthy here, as we impose neutral stratification on the atmosphere later in the model (Section \ref{Section_Response_NS}), that one can argue that the $g\delta\rho$ term in Eq.\eqref{momentum_r_App} is supposed to vanish since it is often referred to as the \textit{buoyancy} term. However, this is not exactly the case since internal gravity waves are not the only source of density fluctuations. In fact, density fluctuations can also result from the planetary-scale compressibility waves, read Lamb modes. As such, the atmosphere can be at the same time neutrally stratified and still feature strong density variations across the horizontal direction. With this subtlety clarified, the tidal perturbations are then expanded in Fourier series of time and longitude, such that the tidal excitation frequency is denoted by $\sigma$. We introduce the reference velocity $v_0$, and we non-dimensionalize all the variables; namely:
\begin{align}\nonumber
    &dz=H dx, \hspace{.4cm} t=\sigma^{-1}\tilde{t},\hspace{.4cm}\boldsymbol{V}=v_0\tilde{\boldsymbol{V}},\hspace{.4cm}\delta p=p_0\tilde{\delta p},\\\nonumber
&\delta\rho=\rho_0\tilde{\delta\rho},\hspace{.3cm}\delta T=T_0\tilde{\delta T},\hspace{.3cm}U=v_0^2\tilde{U},\hspace{.3cm}G=\frac{\sigma}{\Gamma_1}\tilde{G}, \\\label{non_dimensionlaizaion_eqns}
&\nabla_{\rm h}\cdot = R_{\rm p}^{-1}\tilde \nabla\cdot,\hspace{.4cm}\nabla_{\rm h}=R_{\rm p}^{-1}\tilde \nabla,\hspace{.4cm}J=\frac{\sigma g H}{\kappa}\tilde{J}\,,
\end{align}
Under these definitions, the material derivative now reads
\begin{equation}
    \frac{Dy}{Dt}=\sigma y_0\left[\partial_{\tilde{t}}\tilde{\delta y}+ \frac{v_0}{\sigma H}\frac{d\ln y_0}{dx}\tilde{V}_r\right]\,,
\end{equation}
and the calculation variable $\tilde{G}$ becomes
\begin{equation}
    \partial_{\tilde{t}}\tilde{\delta p}+ \frac{v_0}{\sigma H}\frac{d\ln p_0}{dx}\tilde{V}_r = -\tilde{G}.
\end{equation}
We next set $v_0=\sigma R_{\rm p}$ and introduce the ratios of length scale $\eta =R_{\rm p}/H$, and frequency $\alpha=\sigma/\sigma_0$. Allowing for these changes of variables in the governing system of equations, one directly obtains a dimensionless system in the form:
\begin{align}
 & \partial_{\Tilde{t}}\Tilde{V}_\theta -\nu\cos\theta\Tilde{V}_{\varphi} = -\partial_{\theta}\left(\frac{gH}{v_0^2}\Tilde{\delta p} - \Tilde{U}\right)\,, \label{momentum_theta}\\
& \partial_{\Tilde{t}}\Tilde{V}_\varphi +\nu\cos\theta\Tilde{V}_{\theta} = -\frac{1}{\sin\theta}\partial_{\varphi}\left(\frac{gH}{v_0^2}\Tilde{\delta p} - \Tilde{U}\right) \,,\label{momentum_phi}\\
& (\partial_x -1)\Tilde{\delta p} = -\Tilde{\delta \rho} + \frac{v_0^2}{gH}\partial_x\Tilde{U}\,,\label{momentum_r}\\
&\partial_{\Tilde{t}}\Tilde{\delta\rho}+\Tilde{\nabla}\cdot\Tilde{\boldsymbol{V}}=-\eta\left(\partial_x + \frac{d\ln{\rho_0}}{dx}\right)\Tilde{V}_r\,,\label{mass_cons}\\
&\frac{\Tilde{G}}{\Gamma_1}+\partial_{\Tilde{t}}\Tilde{\delta\rho}+\eta \frac{d\ln\rho_0}{dx}\Tilde{V}_r = -\Tilde{J}+ \frac{\delta \Tilde{T}}{\alpha}\,,\label{heat_eqn}\\
&\Tilde{\delta p} = \Tilde{\delta \rho} + \tilde{\delta T}.\label{ideal_gas_law}
\end{align}
Due to the periodic nature of the tidal forcing, the tidal response is Fourier-decomposed in time and longitude, allowing us to write all the physically varying quantities in the form
\begin{equation}\label{Fourier_expansion}
f(x,\theta,\varphi,\tilde{t})=\Re\left\{\sum_m f^{m,\nu}(x,\theta) e^{i( \tilde{t} + m\varphi)}\right\},
\end{equation}
where $m$ is the longitudinal degree. Furthermore, due to the traditional approximation, the decoupling of the vertical and horizontal structures of tidal dynamics allows the  expansion of the Fourier coefficients of Eq. \eqref{Fourier_expansion} into series of Hough functions \citep{hough} describing the horizontal tidal flow; namely:
\begin{equation}\label{Hough_expansion}
    f^{m,\nu}(x,\theta) = \sum_n f_n^{m,\nu}(x)\Theta_n^{m,\nu}(\theta).
\end{equation}
Unless stated otherwise, we denote throughout the paper the coefficients $f_n^{m,\nu}$ by $f_n$ for simplicity. As such, the Laplace tidal equation given in the main text by Eq.\eqref{Laplace_Tidal_equation} is rewritten as:
\begin{equation}
\mathcal{L}\Theta_n= -\Lambda_n\Theta_n,
\end{equation}
where the operator $\mathcal{L}$ is defined as
\begin{align}\label{horizontal_operator}\nonumber
    \mathcal{L} &= \frac{1}{\sin\theta}\frac{d}{d\theta}\left(\frac{\sin\theta}{1-{\nu}^2\cos^2\theta}\frac{d}{d\theta}\right) \\ 
   &- \frac{1}{1-{\nu}^2\cos^2\theta}
    \left(m{\nu}\frac{1+{\nu}^2\cos^2\theta}{1-{\nu}^2\cos^2\theta} + \frac{m^2}{\sin^2\theta}\right).
\end{align}
\section{Retrieving the wave equation of vertical dynamics}\label{VSE_appendix}
The governing system   of  dimensionless equations (\ref{momentum_theta}-\ref{ideal_gas_law}) decouples into equations describing the horizontal structure (\ref{momentum_theta}-\ref{momentum_phi}), and those describing the vertical structure. In the Fourier domain, the latter equations read:
\begin{align}
    &i\tilde{\delta p}_n - \eta\tilde{V}_{r;n} = -\tilde{G}_n,\label{Eq1}\\
    &\!\!\left(\partial_x -1\right)\tilde{\delta p}_n = -\tilde{\delta \rho}_n + \frac{v_0^2}{gH}\partial_x \tilde{U}_n,\label{Eq2}\\
    &\tilde{\delta\rho}_n\!-\!i\eta\left(\!\partial_x\!+ \!\frac{d\ln\rho_0}{dx}\!\right)\!\tilde{V}_{r; n} \!= \!-\mathcal{L}
    \!\left(\frac{gH}{v_0^2}\tilde{\delta p}_n -\tilde{U}_n\!\right)\!,\label{Eq3}\\
    &\frac{\tilde{G}_n}{\Gamma_1}+i\tilde{\delta\rho}_n+\eta\frac{d\ln\rho_0}{dx}\tilde{V}_{r;n} = -\tilde{J}_n + \frac{\tilde{\delta T}_n}{\alpha},\label{Eq4} \\ 
    &\tilde{\delta p}_n=\tilde{\delta \rho}_n+\tilde{\delta T}_n.\label{Eq5}
\end{align}
To recover the wave equation \eqref{wave_equation} from this system, we aim to reduce this system to a single second order partial differential equation in the calculation variable $\tilde{G}_n$. First, we combine Eq. \eqref{Eq1} with the heat transfer equation \eqref{Eq4} to eliminate the vertical component of velocity, $\tilde{V}_{r; n}$. Then we use the ideal gas law \eqref{Eq5} to replace $\tilde{\delta T}_n$ by $\tilde{\delta p}_n$ and $\tilde{\delta \rho}_n$ in the resulting equation. It follows that
\begin{equation}\label{aux11}
\tilde{\delta p}_n\!\left(\!1\!+ \!\frac{d\ln H}{dx\!}-\!\frac{i}{\alpha}\right) - \tilde{\delta \rho}_n\left(\!1\!-\!\frac{i}{\alpha}\right) \!= \frac{i N^2_{\rm B}H}{g}\tilde{G}_n - i\tilde{J}_n.
\end{equation}
Next, we use the hydrostatic equilibrium condition \eqref{Eq2} to express $\tilde{\delta \rho}_n$  in terms of $\tilde{\delta p}_n$ and $\tilde{U}_n$ in the above equation to obtain
\begin{align}\nonumber 
    \partial_x\tilde{\delta p}_n &= i\frac{\alpha H N_{\rm B}^2}{g(\alpha-i)}\tilde{G}_n-\frac{\alpha}{\alpha-i}\frac{d\ln H}{dx}\tilde{\delta p}_n-\frac{i\alpha}{\alpha-i}\tilde{J}_n\\
    &+\frac{v_0^2}{gH}\partial_x\tilde{U}_n.\label{partial_p}
\end{align}
Now our aim is to obtain another first order equation in the calculation variable $\tilde{G}_n$. We start with the hydrostatic equilibrium condition \eqref{Eq2}, and we substitute $\tilde{\delta p}_n$ by its expression \eqref{Eq1}. Then we use the continuity equation \eqref{Eq3} to express $\tilde{\delta\rho}_n$ as a function of $\tilde{V}_{r; n}$ and $\tilde{\delta p}_n$, and we finally eliminate $\tilde{V}_{r; n}$ using equation \eqref{Eq1} to obtain
\begin{equation}\label{partial_G}
    \partial_x \tilde{G}_n = \tilde{G}_n -i\frac{gH}{v_0^2}\mathcal{L}\tilde{\delta p}_n + i\left(\mathcal{L} - \frac{v_0^2}{gH}\partial_x\right)\tilde{U}_n.
\end{equation}
We thus have two first order partial differential equations (\ref{partial_p}\,-\,\ref{partial_G}) of the form
\begin{align}\label{first_order_G}
    &\partial_x\tilde{G}_n = A_1\tilde{G}_n + B_1\tilde{\delta p}_n + C_1,\\
    &\partial_x\tilde{\delta p}_n = A_2\tilde{G}_n + B_2\tilde{\delta p}_n + C_2.
\end{align}
Taking the derivative of Eq.\eqref{first_order_G}, it is straightforward to obtain a second order equation of the form:
\begin{equation}\label{2nd_order_G}
    \left[\partial_x^2 + A\partial_x + B\right]\tilde{G}_n = C_n,
\end{equation}
where
\begin{align}\label{coeff_A}
    &A(x) = -1 + \frac{i}{\alpha-i}\frac{d\ln H}{dx},\\\label{coeff_B}
    &B(x)=-\frac{\alpha}{\alpha-i}\left[\frac{i}{\alpha}\frac{d\ln H}{dx} + \frac{H^2N_{\rm B}^2}{v_0^2}\mathcal{L}\right],\\
    &C_n(x) = -\frac{gH\alpha}{v_0^2(\alpha-i)}\mathcal{L}\tilde{J}_n-\Bigg[i\frac{d}{dx}\left(\frac{v_0^2}{gH}\frac{d}{dx}\right)\\\label{coeff_C}
    &\hspace{.8cm}+ \frac{1}{\alpha-i}\frac{d\ln H }{dx}\left(\mathcal{L}-\frac{v_0^2}{gH}\frac{d}{dx}\right)\Bigg]\tilde{U}_n.
\end{align}
Next, we implement a change of variable of the form $\tilde{G}_n(x)=\Phi(x)\Psi_n(x)$ to write Eq. \eqref{2nd_order_G} in the standard form of wave equations. We find that the first order term in Eq. \eqref{2nd_order_G} cancels when
\begin{equation}
    2\frac{d\Phi(x)}{dx}+ A(x)\Phi(x) = 0,
\end{equation}
which is satisfied by
\begin{align}
    \Phi(x) &= \exp\left(-\frac{1}{2}\int_0^x A(x^\prime)dx^\prime\right).
\end{align}
This gives us the form of $\Phi(x)$ defined in Eq.\eqref{Eq_Phi_x}, and consequently the wave equation \eqref{wave_equation}, with a vertical wavenumber $\hat{k}$  defined as
\begin{align}\nonumber
    \hat{k}_n^2&=-\frac{1}{4}\left(A^2+2\frac{dA}{dx}-4B\right)\\\nonumber
    &=-\frac{1}{4}\Bigg\{\!\!\!\left(\!1\!-\!\frac{i}{\alpha-i}\frac{d\ln H}{dx}\right)^2\!\!\!\!+2\frac{d}{dx}\!\!\left(\frac{i}{\alpha-i}\frac{d\ln H}{dx}\right)\\
    &-\frac{4\alpha}{\alpha-i}\!\left[\frac{gHN_{\rm B}^2\Lambda_n}{v_0^2}-\!\frac{i}{\alpha}\frac{d\ln H}{dx} \right]\Bigg\}.\label{wavenumber1}
\end{align}
With the definitions of the dimensionless control parameters given in Table \ref{tab:control_param}, the wave number $\hat{k}_n$ (Eq.\ref{wavenumber1}) and the forcing term (Eq.\ref{coeff_C}) can be rewritten in the form given in the main text by Eqs.(\ref{wavenumber2}) and (\ref{C_equation}) respectively.

\section{Solutions of the vertical profiles of tidal variables}\label{polarizations_appendix}
In this section we provide the solutions to the vertical profiles of the tidally varying scalar fields of pressure, density, and temperature, and the velocity vector field, given the solution of the wave equation $\Psi(x)_n$, or equivalently the variable $\tilde{G}_n$.

First, the expression of pressure variations is readily obtained from Eq.\eqref{partial_G}, which for a given Hough mode and using the definition of $\beta$ (Table \ref{tab:control_param}) reads
\begin{equation}\label{delta_p_solution}
    \tilde{\delta p}_n = \frac{1}{i\beta\Lambda_n}\left(\frac{d\tilde{G}_n}{dx}-\tilde{G}_n\right) + \frac{1}{\beta}\left(1+\frac{1}{\beta\Lambda_n}\frac{d}{dx}\right)\tilde{U}_n.
\end{equation}
Given $\tilde{\delta p}_n(x)$, the horizontal component of the velocity field is straightforwardly obtained via \citep[e.g., Eqs.(25-26) of][]{auclair2017rotation}:
\begin{align}
    \tilde{V}_{\theta; n} &= \frac{\partial_\theta+m\nu \cot\theta}{1-\nu^2\cos^2\theta}\left(\frac{gH}{v_0^2}\tilde{\delta p}_n-\tilde{U}_n\right),\\
     \tilde{V}_{\varphi; n} &= -\frac{\nu\cos\theta\partial_\theta+{m}{\sin^{-1}\theta}}{1-\nu^2\cos^2\theta}\left(\frac{gH}{v_0^2}\tilde{\delta p}_n-\tilde{U}_n\right),
\end{align}
which gives
\begin{align}
      &\tilde{V}_{\theta; n} &=\frac{1}{\Lambda_n}\left(\frac{d\tilde{G}_n}{dx}-\tilde{G}_n + i\beta^{-1} \frac{d\tilde{U}_n}{dx}\right),\\
      &\tilde{V}_{\varphi; n} &=\frac{i}{\Lambda_n}\left(\frac{d\tilde{G}_n}{dx}-\tilde{G}_n + i\beta^{-1} \frac{d\tilde{U}_n}{dx}\right).
\end{align}
The vertical component of the velocity is established from equation \eqref{Eq1}, replacing $\tilde{\delta p}_n$ by its solution \eqref{delta_p_solution}:
\begin{align}\nonumber
    \tilde{V}_{r; n} &= \frac{1}{\eta\beta\Lambda_n}\left[\frac{d\tilde{G}_n}{dx}+\left(\beta\Lambda_n-1\right)\tilde{G}_n\right]\\
    &+\frac{i}{\eta\beta}
\left(1+\frac{1}{\beta\Lambda_n}\frac{d}{dx}\right)\tilde{U}_n.\label{vertical_velocity}
\end{align}
Finally, the solution for the density perturbation is obtained by replacing the solution of $\tilde{\delta p}_n$ \eqref{delta_p_solution} in equation \eqref{aux11} to get:
\begin{align}\nonumber
    \tilde{\delta \rho} _n&= \frac{1}{i\beta\Lambda_n}\left(1+\frac{\kappa\alpha(\gamma-1)}{\alpha-i}\right)\Bigg[\frac{d\tilde{G}_n}{dx} +\\\nonumber
&\left(\frac{\alpha\beta\gamma\kappa{\Lambda}}{\alpha-i+\kappa\alpha(\gamma-1)}-1\right)\tilde{G}_n\Bigg]+\frac{i\alpha\tilde{J}_n}{\alpha-i}\\
&+ \frac{1}{\beta}\left(1+\frac{\kappa\alpha(\gamma-1)}{\alpha-i}\right)\left(1+\frac{1}{\beta\Lambda_n}\frac{d}{dx}\right)\tilde{U}_n,
\end{align}
and the vertical profile of temperature is then readily obtained by the ideal gas law \eqref{Eq5}.

\section{Thermal exchange mechanisms and the total propagating flux}\label{appendix_thermal_Pbl}

\begin{figure}
\centering
\includegraphics[width=.4\textwidth]{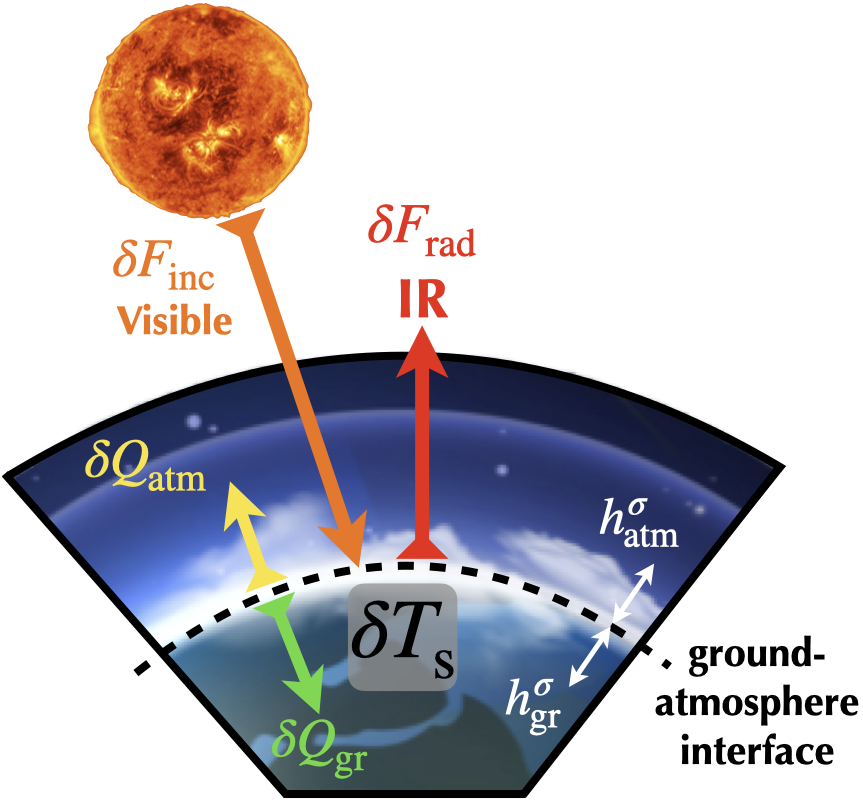}
    \caption{\RVA{A schematic of the bichromatic radiative and diffusive flux exchange mechanisms at the surface interface between the lower troposphere and the ground. The power balance between the fluxes in given by Eq. \eqref{power_budget_1}. } }
    \label{PBL_Schematic}
\end{figure}

We define here explicitly the thermal exchange mechanisms used to establish the thermal energy budget at the surface and to construct the thermal forcing profile given by Eq.\eqref{delta_f_tot} of Section \ref{Sec_thermal_forcing}. 

 As mentioned in the main text, the incident stellar flux, $\delta F_{\rm inc}$ generates a variation $\delta T_{\rm s}$ in the surface temperature $T_{\rm s}$. We denote the proportionality  by the complex transfer function $ \mathcal{B}_{\rm gr}^{\sigma}$ such that $\delta T_{\rm s}=\mathcal{B}_{\rm gr}^{\sigma}\delta F_{\rm inc}$. The  variation $\delta T_{\rm s}$ induces radiative emission with magnitude $\delta F_{\rm rad}$. A  fraction of the incident power, $\delta Q_{\rm gr}$,  is then transmitted  to  the  ground  by  thermal  conduction,  and  another fraction, $\delta Q_{\rm atm}$, is transmitted to the atmosphere through turbulent  thermal  diffusion \RVA{(see Figure \ref{PBL_Schematic})}. \RVR{Finally, as the  atmosphere  is  heated, it  undergoes radiative  cooling, in a similar fashion  to the surface.  We denote by $\delta F_{\rm atm}$ the latter atmospheric flux impinging upon the  surface.}{} 

Having isolated these thermal mechanisms, the total power balance of the thermal perturbation at the  surface interface is expressed as\footnote{we ignore latent heat exchanges associated with changes of state.}:
\begin{equation}\label{power_budget_1}
     \delta F_{\rm inc}-\delta F_{\rm rad}  =  \delta Q_{\rm gr} + \delta Q_{\rm atm} .
\end{equation}
To define each of these terms explicitly, we start with the surface radiative emission term $\delta F_{\rm rad}$, which can be obtained by differentiating the Stefan-Boltzmann law as a function of the surface temperature assuming that the planetary surface radiates as a black body. Namely we have
\begin{equation}
    \delta F_{\rm rad} = 4\sigma_{\rm SB}T_{\rm s}^3\delta T_{\rm s},
\end{equation}
with $\sigma_{\rm SB}=5.670373\times10^{-8}$ W\,m$^{-2}$\,K$^{-4}$ being the Stephan-Boltzmann constant \citep{tiesinga2021codata}.  \RVR{Without any loss of generality,  and justified by the small magnitude of the atmospheric emission towards the surface, one can assume that $\delta F_{\rm atm}$ is also proportional to $\delta T _{\rm s}$, and combine the two radiative terms $\delta F_{\rm rad}$ and $\delta F_{\rm atm}$ into a single term with an effective emissivity. This assumption follows from: \textit{i)} the fact that the bulk atmosphere responsible for the greenhouse effect is formed by the lowermost atmospheric layers where temperature oscillations are close to the surface temperature oscillations; \textit{ii)} the fact that energy given to the atmosphere is a fraction of $\delta F_{\rm rad}$, and the flux $\delta F_{\rm atm}$ is a fraction of this input energy, which implies it should have the same form up to a scaling factor.}{}

Next, the terms associated with heat diffusion at the boundary interface, $ \delta Q_{\rm gr}$ and $ \delta Q_{\rm atm}$, are derived using the gradient flux theory \citep{garratt1994atmospheric}:
\begin{equation}
    \delta Q_{\rm gr}(t) =\frac{k_{\rm gr}}{H}\frac{\partial \delta T}{\partial x}\Bigg|_{x=0^{-}},
\end{equation}
\begin{equation}
    \delta Q_{\rm atm}(t) =-\frac{k_{\rm atm}}{H}\frac{\partial \delta T}{\partial x}\Bigg|_{x=0^{+}}.
\end{equation}
\RVA{Here, $k_{\rm gr}$ is the thermal conductivity of the ground, and $k_{\rm atm}$ is its atmospheric analogue. We also denote by $x=0^-$ and $x=0^+$ the vicinity below and above the surface interface, respectively.} Temperature variations near the surface can be traced by the heat transport equation, which in the frequency domain reads as \citep[e.g.,][]{chapman1969atmospheric}
\begin{equation}
    i\sigma\delta T = \frac{K_{\rm gr}}{H^2}\frac{\partial^2\delta T}{\partial x^2}\,\,\,\,\,{\rm for}\,\, x<0\,,
\end{equation}
\begin{equation}
    i\sigma\delta T = \frac{K_{\rm atm}}{H^2}\frac{\partial^2\delta T}{\partial x^2}\,\,\,\,\,{\rm for}\,\, x>0.
\end{equation}
\RVA{with $K_{\rm gr}=k_{\rm gr}/(\rho_0(0^-)C_{\rm gr})$ and $K_{\rm atm}=k_{\rm atm}/(\rho_0(0^+)C_{\rm p})$ being the thermal diffusivities near the surface in the ground and the atmosphere respectively, $C_{\rm gr}$ and $C_{\rm p}$ being the respective thermal capacities per unit mass, while $\rho_0(0^-)$ and $\rho_0(0^+)$ are the densities in the vicinity of the surface interface.} These equations have solutions of the form:
\begin{equation}\label{solution_delta_tx-}
    \delta T(x) = \delta T_{\rm s} \exp \left\{ (1+is)x/h_{\rm gr}^\sigma\right\},\,\,\,\,{\rm for}\,\, x\leq0,
\end{equation}
\begin{equation}\label{solution_delta_tx+}
    \delta T(x) = \delta T_{\rm s} \exp \left\{-(1+is)x/h_{\rm atm}^\sigma\right\},\,\,\,\,{\rm for}\,\, x>0.
\end{equation}
Here, \RVA{we remind the reader that $s=\text{sgn}({\sigma})$}, and we denote by $h_{\rm gr}^\sigma$ and $h_{\rm atm}^\sigma$ the skin thicknesses of heat transport by thermal diffusion in the ground and the atmosphere defined as
\begin{equation}
    h_{\rm gr}^\sigma = \frac{1}{H}
\frac{2K_{\rm gr}}{|\sigma|}\hspace{.6cm}{\rm and}\hspace{.6cm}h_{\rm atm}^\sigma = \frac{1}{H}
\frac{2K_{\rm atm}}{|\sigma|}.
\end{equation}
To obtain the explicit form of $ \mathcal{B}_{\rm gr}^{\sigma}$, we first rewrite the power budget balance of Eq.\eqref{power_budget_1} as
\begin{equation}\label{power_budget_2}
     \delta F_{\rm inc} - \frac{k_{\rm gr}}{H}\frac{\partial \delta T}{\partial x}\Bigg|_{x=0^{-}} \!\!\!\!\!\!\!\!\!= -\frac{k_{\rm atm}}{H}\frac{\partial \delta T}{\partial x}\Bigg|_{x=0^{+}}  \!\!\!\!\!\!\!\!\!+4\sigma_{\rm SB}T_{\rm s}^3\delta T_{\rm s}\,,
\end{equation}
and we substitute the solutions of $\delta T(x)$ of Eq.~\eqref{solution_delta_tx-} to obtain 
\begin{equation}\label{eq_bgr}
    \mathcal{B}_{\rm gr}^\sigma = \left\{ \left(4\sigma_{\rm SB}T_{\rm s}^3\right) \left[ 1+ (1+si)\sqrt{\tau_{\rm bl}|\sigma|}\right]\right\}^{-1}.
\end{equation}
Here we have defined the boundary layer timescale $\tau_{\rm bl}$ as
\begin{equation}\label{eq_tau_bl}
    \tau_{\rm bl}= \frac{1}{2}\left( \frac{I_{\rm gr}+I_{\rm atm}}{4\sigma_{\rm SB}T_{\rm s}^3}\right)^2\,,
\end{equation}
where the functions $I_{\rm gr}$ and $I_{\rm atm}$ denote, respectively, the thermal inertias of the ground and the atmospheric surface layers, and are specifically given by:
\begin{align}\nonumber
    I_{\rm gr}=& \rho_0(0^-)C_{\rm gr}\sqrt{K_{\rm gr}}\,, \\
 I_{\rm atm}=&\rho_0(0^+)C_{\rm p}\sqrt{K_{\rm atm}}\,.\,\label{thermal_inertias}
\end{align}
Finally, with $\delta Q_{\rm gr}$ written as \begin{equation}\label{eq_Qge}
    \delta Q_{\rm gr} = \frac{k_{\rm gr}}{\sqrt{2K_{\rm gr}}}\sqrt{|\sigma|}\mathcal{B}_{\rm gr}^\sigma(1+si) \delta F_{\rm inc},
\end{equation}
we are fully geared to define the total  propagating flux feeding the atmosphere, $\delta F_{\rm tot}$, as
\begin{equation}\label{eq_delta_ftot}
    \delta F_{\rm tot}=  \delta F_{\rm inc} - \delta Q_{\rm gr},
\end{equation} 
from which we retrieve Eq.\eqref{delta_f_tot} of the main text. \RVA{As for the numerical values of the material properties considered in the formalism above, we first set the gas heat capacity to its nominal value $C_{\rm p}=1005$~J/kg$\cdot$K \citep{hilsenrath1955tables}. The volumetric heat capacity in the ground $\rho_0(0^-)C_{\rm gr}$ varies between the land and the oceans, but since we're considering an effective value over a rocky planet, we adopt the typical value of $2.5\times10^6$~J\,m$^3$\,K$^{-1}$ \citep{van1963periodic}. The value for the density $\rho_0$ is dependent on the planetary setting understudy and is computed via $\rho_0=p_0/gH$. The values for ground thermal conductivities and diffusivities vary among rock types and water content taking typical values of $k_{\rm gr}\sim 0.5$~--~$4$~W\,m$^{-1}$\,K$^{-1}$ and $K_{\rm gr}$~$\sim$~$ 10^{-6}$~--~$10^{-8}$~m$^2$\,s$^{-1}$\citep[e.g.,][]{andujar2016ground,arkhangelskaya2018estimating}. In contrast, for the atmosphere, the turbulent eddy diffusivity $K_{\rm atm}$ is known to have a strong altitude dependence, but the profile of this dependence varies significantly in the literature between linear, exponentially decaying, and higher order polynomial profiles with several inflection points, with near surface estimates ranging between $0.1$ and $100$~m$^2$~s$^{-1}$ \citep[e.g.,][]{bernard1962theorie,madsen1977realistic,nieuwstadt1983solution,holtslag1993local,berger1998baroclinic}. Specifically, \citet{bernard1962theorie}
 computes an average effective value over continental areas $\overline{K}_{\rm atm}=3.6$~m$^2$~s$^{-1}$. This variability of the ground and atmospheric thermal parameters propagates to our model free parameter $\tau_{\rm bl}$, or equivalently the frequency $\sigma_{\rm bl}$, via the thermal ground and atmospheric thermal inertias defined in Eq. \eqref{thermal_inertias}. Thus in sweeping the limits of $\sigma_{\rm bl}$ in Figure \ref{Lamb_asymmetry}, we are moving between strong and weak thermal inertia limits by self-consistently varying the values of the thermal properties. For instance, the value of $\sigma_{\rm bl}=10^{-5}$~s$^{-1}$, for which the accelerating torque at the Lamb resonance completely vanishes, corresponds to  
 $k_{\rm gr}= 1.25$~W\,m$^{-1}$\,K$^{-1}$, $K_{\rm gr}= 5\times10^{-7}$~m$^2$\,s$^{-1}$, and $K_{\rm atm}=\overline{K}_{\rm atm}=3.6$~m$^2$~s$^{-1}$.}

\section{The quadrupolar component of the incident stellar flux}\label{quadrupolar_forcing_appendix}
As we are interested in the semi-diurnal tidal response, we decompose here the incoming stellar flux harmonically to isolate the quadrupolar component. We start by expressing the day-night periodically varying incident flux $\delta F_{\rm inc}$ (denoted hereafter by $F$ for convenience) as:
\begin{equation}
  F=  \begin{cases}
      F_\star\cos\Theta, & \text{for}\ 0\leq\Theta\leq\pi/2 \\
      0, & \text{otherwise}\,,
    \end{cases}
  \end{equation}
where $\Theta$ is the zenith angle. To obtain the quadrupolar component of the thermal forcing we first expand $F$ in Legendre Polynomials:
\begin{equation}
     F(\Theta) = \sum_{l=0}^{\infty} F_lP_l(\cos\Theta),
\end{equation}
where the expansion coefficients are given by
\begin{align}\nonumber
     F_l &= \frac{2l+1}{2}\int_0^{\pi}  F(\Theta)P_l(\cos\Theta)\sin\Theta d\Theta \,,\\
     &=  \frac{2l+1}{2} F_\star\int_0^{\pi/2}  P_l(\cos\Theta)\sin\Theta\cos\Theta d\Theta.
\end{align}
Next, using the addition theorem \citep[e.g.,][]{abramowitz1988handbook}, we write the Legendre Polynomials as series of
spherical harmonics:
\begin{equation}
    P_l(\cos\Theta) =\frac{4\pi}{2l+1}\sum_{m=-l}^{m=l}Y_{lm}(\theta,\varphi)Y_{lm}^*(\theta_{\rm S}, \varphi_{\rm S})\,,
\end{equation}
where the asterisk corresponds to complex conjugation, and the coordinates $(\theta_{\rm S}, \varphi_{\rm S})= \left[\pi/2, (\Omega-n_\star)t\right]$ define the position of the star when ignoring planetary obliquity. We thus have:
\begin{equation}\label{F_aux1}
    F = \sum_{l=0}^{\infty} \sum_{m=-l}^{m=l}C_l F_lY_{lm}(\theta,\varphi)Y_{lm}^*(\theta_{\rm S}, \varphi_{\rm S})\,,
\end{equation}
where we have defined $C_l={4\pi}/(2l+1) $. Using the definition of spherical harmonics 
\begin{equation}\label{Spherical_Harmonics_definition}
    Y_{lm}(\theta,\varphi) = N_{lm} P_{lm}(\cos\theta)e^{im\varphi},
\end{equation}
with 
\begin{equation}
    N_{lm} = \sqrt{\frac{(2l+1)(l-m)!}{4\pi(l+m)!}}\,,
\end{equation}
and the associated Legendre functions
\begin{equation}
    P_{lm}(\cos\theta)= (-1)^m(1-\cos^2\theta)^\frac{m}{2} \frac{d^m}{dx^m}P_l(\cos\theta),
\end{equation}
we re-write Eq.\eqref{F_aux1} as:
\begin{align}\nonumber
      &F=  \sum_{l=0}^{\infty} \sum_{m=-l}^{m=l}\gamma_{lm}(\theta)\cos\left[m(\sigma_{11} t+\varphi)\right] \\
      &+ i \sum_{l=1}^{\infty} \sum_{m=1}^{m=l}\sin\left[m(\sigma_{11} t+\varphi)\right]\left[\gamma_{lm}(\theta)- \gamma_{l-m}(\theta)\right]\,,
\end{align}
where $\gamma_{lm}(\theta)= C_lF_lN_{lm}^2 P_{lm}(\cos\theta)P_{lm}(0)$, and $\sigma_{11} = \Omega-n_\star$.  Next we note that
\begin{align}\nonumber
    &\gamma_{lm}(\theta)- \gamma_{l-m}(\theta)= C_lF_l\Big[N_{lm}^2P_{lm}(\cos\theta)P_{lm}(0) \\\nonumber
   &- N_{l-m}^2P_{l-m}(\cos\theta)P_{l-m}(0)\Big]=0,
\end{align}
by virtue of the equality
\begin{equation}
    P_{l-m}=\frac{(l-m)!}{(l+m)!}P_{lm}.
\end{equation}
Finally, $F$ is written as the real part of the thermal forcing function in the Fourier domain:
\begin{equation}
    F= \Re\left\{\sum_{l=0}^{\infty}\sum_{m=0}^{l}(2-\delta_{0m})\gamma_{lm}(\theta)e^{im(\sigma_{11} t+\varphi)}\right\},
\end{equation}
from which we retrieve the quadrupolar component ($l=m=2)$ of the forcing in the form
\begin{equation}
   {F}_{22}(\theta,\varphi,t) = \frac{5}{64}F_\star P_{22}(\cos\theta)e^{i(\sigma_{22}t+2\varphi)},
\end{equation}
or in spherical harmonics as
\begin{equation}\label{App_eq_quadrupolar_flux}
    {F}_{22}(\theta,\varphi,t) = \frac{\sqrt{30\pi}}{16}F_\star Y_{22}(\theta,\varphi)e^{i\sigma_{22}t}, 
\end{equation}
where $\sigma_{22}= 2(\Omega- n_\star)$ is the semi-diurnal tidal forcing frequency. 
\section{Energy carried by tidal waves}\label{appendix_energy_flow}
Essential to the closed-form solution we provide for the tidal response is the boundary condition imposed against tidal waves at the uppermost layer of the atmosphere. There we enforce a non-divergence condition on the tidal flow, namely, the energy of tidal flows $\mathcal{W}$ should be bounded as $x\rightarrow\infty$. It follows from \cite{Wilkes49} that the energy flow associated with the tidal $(n,m,\nu)$-mode across the vertical direction is given by
\begin{equation}
    \mathcal{W} = \frac{1}{2}p_0v_0\Re\left\{\tilde{\delta p}\tilde{V}_r^*\right\}.
\end{equation}
By substituting $\tilde{V}_r$ by its expression in Eq.\eqref{Eq1}, we notice that
\begin{equation}
    \Re\left\{\tilde{\delta p}\tilde{V}_r^*\right\} = \eta^{-1}\Re\left\{\tilde{\delta p}\tilde{G}^*\right\}.
\end{equation}
Therefore, using Eq.\eqref{pressure_anomaly}, we define the tidal energy flow as a function of the wave equation solution as
\begin{align}\nonumber
    \mathcal{W} &= \frac{p_0v_0}{2\eta\beta\Lambda}\Im\left\{\frac{d\tilde{G}}{dx}\tilde{G}^*\right\}\\
     &+ \frac{p_0v_0}{2\eta\beta}\Re\left\{\left(1+ \frac{1}{\beta\Lambda}\frac{d}{dx}\right)\tilde{U}\tilde{G}^*\right\}.
\end{align}
Ignoring the gravitational tidal potential $\tilde{U}$, and using the variables $\Psi$ and $\Phi$ of the wave equation in the main text, we re-write the energy flow as
\begin{equation}\nonumber
     \mathcal{W}= \frac{p_0v_0}{2\eta\beta\Lambda}\Im\left\{\frac{d\Psi}{dx}\Psi^* - \frac{A}{2}|\Psi|^2\right\}|\Phi|^2,
\end{equation}
which scales as $|\Psi|^2 |\Phi|^2$ as we discuss in the main text. 
\section{The tidal torque exerted about the spin axis}\label{Appendix_Torque}
At first glance, the definitions of the atmospheric tidal torque, as a function of the pressure anomaly variation, present in various works of the literature may seem inconsistent [see eg., Eq.(27) of \cite{dobrovolskis1980atmospheric}; Eq.(3) of \cite{zahnle1987constant}; Eq.(4) of \cite{auclair2019generic}]. Upon a more careful inspection, one can relate the evident mismatch to the adopted convention in defining the pressure anomaly, which when accounted for reveals the overlap between the different expressions.  Here we recover the expression of the tidal torque that we use in Eq.\eqref{Equation_Torque}. 

The instantaneous torque exerted by tidal force about the spin axis of a planet of volume $\mathcal{V}$ can be expressed as 
\begin{equation}
\mathcal{T}=\int_\mathcal{V}\partial_\varphi U_{\rm T}d\mu,
\end{equation}
where $U_{\rm T}(\Vec{r},t)$ represents the tidal potential, $\Vec{r}$ being the separation between the tidal players, and $d\mu$ being  an infinitesimal mass parcel of the tidally induced mass redistribution. i.e. the tidal bulge. \RVA{Under the hydrostatic approximation, where the pressure and gravity forces are balanced, the distribution of mass can be quantified by the pressure variation instead of the density variation \citep[e.g.,][]{leconte2015asynchronous}. Furthermore, since the atmospheric layer is relatively thin, the mass distribution can be further quantified by the surface pressure anomaly. In the case of a thick atmospheric layer, however, one should integrate the mass distribution over the full radial extent of the layer. That said, the torque, averaged over an infinite time period $P$, is written as} 
\begin{equation}\label{Eq_App_Torque_1}
    \mathcal{T}= \lim_{P\rightarrow+\infty}\frac{1}{P}\int_0^P\!\!\int_\mathcal{S}\partial_\varphi U_{\rm T} \frac{R_{\rm p}^2 \delta p_{\rm s}}{g} d{S}dt,
\end{equation}
where $\mathcal{S}$ is a sphere of unit radius, and $d{S}=\sin\theta d\theta d\varphi$. Due to the periodic nature of the forcing and the response, the tidal potential and the pressure anomaly are expanded in Fourier series with $\sigma$ being the tidal frequency such that
\begin{align}
   &U_{\rm T}(\Vec{r},t)= \Re \left\{ \sum_\sigma U_{\rm T}^\sigma(\Vec{r})e^{i\sigma t}\right\},\\
   &\delta p_{\rm s}(\theta,\varphi,t)= \Re \left\{ \sum_\sigma \delta p_{\rm s}^\sigma(\theta,\varphi)e^{i\sigma t}\right\}.
\end{align}
As such, the torque is rewritten as
\begin{equation}\label{Eq_App_Torque_2}
    \mathcal{T}=\frac{R_{\rm p}^2}{2g}\sum_\sigma\Re\left\{\int_\mathcal{S}{\partial_\varphi^* U_{\rm T}^{\sigma *}}\delta p_{\rm s}^\sigma dS \right\}.
\end{equation}
The Fourier coefficients of the tidal potential and the pressure anomaly are further expanded in spherical harmonics (Eq.\ref{Spherical_Harmonics_definition}):
\begin{align}
     &U_{\rm T}^\sigma(\Vec{r})=  \sum_{l=2}^{+\infty}\sum_{m=-l}^{l} U_{{\rm T};l}^{m,\sigma}(r)Y_{l,m}(\theta,\varphi),\\
   &\delta p_{\rm s}^\sigma(\theta,\varphi)=   \sum_{l=2}^{+\infty}\sum_{m=-l}^{l}\delta p_{{\rm s};l}^{m,\sigma}Y_{l,m}(\theta,\varphi).
\end{align}
Substituting these expansions in Eq.\eqref{Eq_App_Torque_2} and using the orthogonality property of spherical harmonics, we obtain:
\begin{equation}\nonumber
       \mathcal{T}=\frac{R_{\rm p}^2}{2g}\sum_\sigma\sum_{l=2}^{+\infty}\sum_{m=-l}^{l}m\Im\left\{{U_{{\rm T};l}^{m,\sigma}}\delta p_{{\rm s};l}^{m,\sigma}\right\}.
\end{equation}
Considering that the harmonic terms of the tidal potential are components of a series in the ratio $R_{\rm p}/a_{\rm p}$, it suffices to truncate the series at the quadrupolar order ($l=m=2)$ to describe the semidiurnal response since $R_{\rm p}\ll a_{\rm p}.$ As such, noting that $g=GM_{\rm p}/R_{\rm p}^2$, and that the quadrupolar component of the potential is given by \citep[e.g.,][]{ogilvie2014tidal}:
\begin{equation}
    U_{{\rm T};2}^{2,\sigma} = \sqrt{\frac{6\pi}{5}}\frac{G M_{\star}}{a_{\rm p}}\left(\frac{R_{\rm p}}{a_{\rm p}}\right)^2, 
\end{equation}
we straightforwardly obtain the expression of the torque in Eq.\eqref{Equation_Torque} \citep[see also][]{leconte2015asynchronous,auclair2019generic}:
\begin{equation}
    \mathcal{T}=\sqrt{\frac{6\pi}{5}}\frac{M_{\star}}{M_{\rm p}}\frac{R_{\rm p}^6}{a_{\rm p}^3}\Im\{\delta p_{{\rm s};2}^{2,\sigma}\}.
\end{equation}

\section{The functional form of the tidal response}\label{App_functional_Form}

\RVA{Our goal in this sections is to provide a more elaborate description of the functional form of the tidal solution obtained from our model. Namely, while Eq.~\eqref{imag_pressure_anomaly} quantifies the imaginary part of the surface pressure anomaly -- as required for computing the tidal torque -- expressed in physical parameters, here we provide the full complex solution in a different form that facilitates future model modifications.
First, the obtained expression of the complex surface pressure anomaly, from which we extract the imaginary part given by Eq.~\eqref{imag_pressure_anomaly}, is given by}
\begin{align}\nonumber
    \tilde{\delta p}_{\rm s}=& \frac{\alpha_{\rm A}\delta F_{22}}{p_s}\frac{\kappa g \Lambda_2}{R_{\rm p}^2 \sigma^3}\frac{\left[\mathcal{X}_{\rm R} + \mathcal{Y}_{\rm R}\alpha + i(\mathcal{X}\alpha+\mathcal{Y})\right]\alpha}{\left(1+2\zeta + 2\zeta^2\right)}\\
    & \times \left[(\kappa-\beta\Lambda_2+1)^2+\alpha^2(\beta\Lambda_2-1)^2\right]^{-1},
\end{align}
\RVA{where $\mathcal{X}$ and $\mathcal{Y}$ are defined by Eq.~\eqref{Chi_Upsilon}, and $\mathcal{X}_{\rm R}$ and $\mathcal{Y}_{\rm R}$ are defined by }
\begin{align}\nonumber
     \mathcal{X}_{\rm R}&= (\kappa\!-\!\beta\Lambda_2\! +\!1)\!\left[2\zeta^2(1-\mu_{\rm gr})\!+\!\zeta(2\!-\!\mu_{\rm gr})+1\right],\\
     \mathcal{Y}_{\rm R}&= s\mu_{\rm gr}\zeta(\beta\Lambda_2 -1).
\end{align}
\RVA{Next, we reformulate this complex solution as
\begin{equation}\label{p_funct_1}
    \tilde{\delta p}_{\rm s}= K \left(\frac{\sigma_{\rm L}}{\sigma}\right)^3 \frac{\mathcal{G}(\sigma)}{\alpha^{-1}\left(\kappa+1-\beta\Lambda_2\right) + i\left(\beta\Lambda_2-1\right)}\,, 
\end{equation}
where we introduced $\sigma_{\rm L}= \sigma_{\rm w}\sqrt{\Lambda_2}$\,, the frequency which marks the peak of the modulus of the resonant pressure variation in the adiabatic limit, the transfer function of the thermal forcing $\mathcal{G}(\sigma)$ defined as
\begin{equation}
    \mathcal{G}(\sigma) = \frac{2\zeta^2(1-\mu_{\rm gr})+\zeta(2-\mu_{\rm gr})+1 - is\mu_{\rm gr}\zeta}{1+2\zeta+2\zeta^2}\,,
\end{equation}
and the dimensionless constant factor $K$ defined as
\begin{equation}\label{Factor_K}
    K = \frac{\alpha_{\rm A}\delta F_{22}}{p_s}\frac{\kappa g \Lambda_2}{R_{\rm p}^2 \sigma_{\rm L}^3}\,.
\end{equation}
The transfer function $\mathcal{G}$ thus describes the delayed forcing of the atmosphere due to thermal inertia near the planetary surface. Therefore, this function reduces to $\mathcal{G} \left( \sigma \right) = 1$ in the case of direct absorption of the incident flux and the synchronous response of the surface (namely, in the limit of $\zeta = \mu_{\rm gr} = 0$, which is the limit with which we describe the Earth in the main text). With $\mathcal{G}(\sigma)$ defined as such, the complex solution is structured in a convenient form where the effects of atmospheric dynamics and the thermal forcing are clearly separated. This facilitates future modifications of the model if one intends, for instance, to replace the thermal forcing function by another.} 

\begin{figure}
\centering
\includegraphics[width=.48\textwidth]{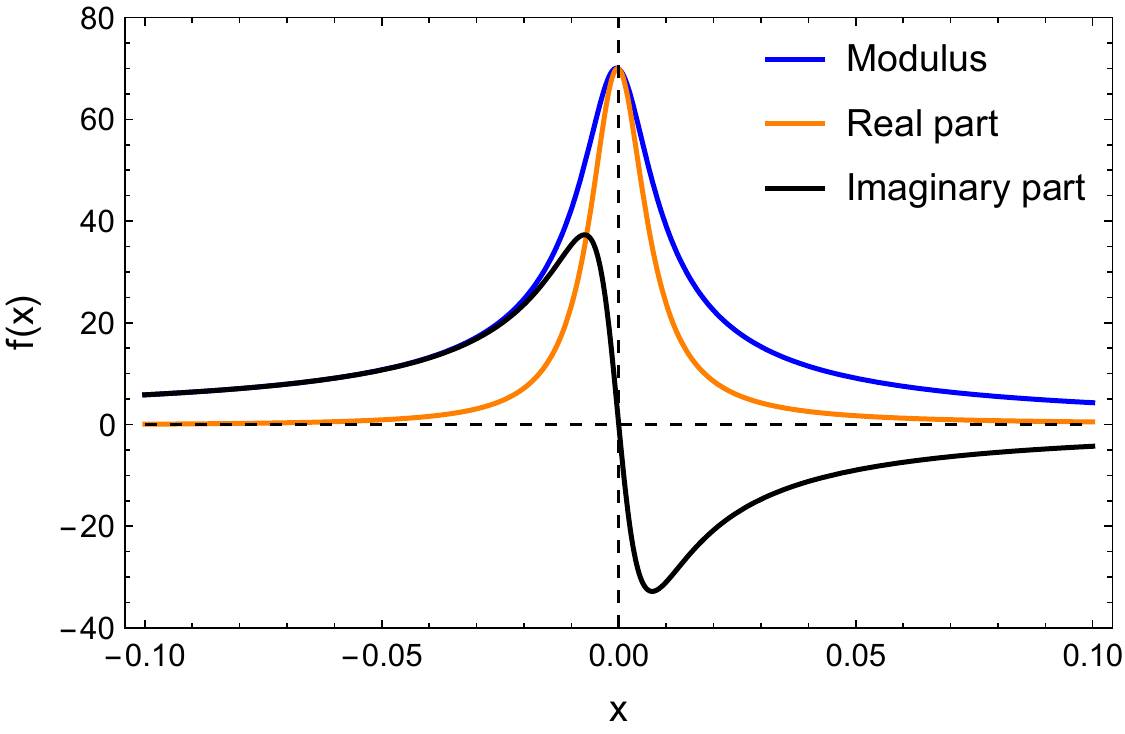}
    \caption{\RVA{A sample of the tidal response, plotted in terms of modulus, real part, and imaginary part of the surface pressure anomaly, all given by Eqs.~\eqref{p_func_form_fin}. The function $f(x)$ on the y-axis indicates the functional form of the pressure variation, that is, the pressure function divided by the constant factor K. The x-axis is centered at the position of the Lamb resonance. Notable is the difference in the resonant amplification of the response between the real part and the imaginary part. } }
    \label{pressure_components}
\end{figure}

\RVA{Next we define the normalized frequency $X=~\sigma/\sigma_{\rm L}$ and the dissipation parameter $\varepsilon=~\sigma_0/\sigma_{\rm L}\ll 1$, and we rewrite Eq.~\eqref{p_funct_1} as
\begin{equation}
    \tilde{\delta p}_{\rm s}= K \frac{\mathcal{G}(X)}{\varepsilon\left[(\kappa+1)X^2-1\right]+iX\left(X^2-1\right)}\,. 
\end{equation} }
\RVA{The modulus, real part, and the imaginary part of this complex surface pressure oscillation are then straightforwardly deduced as:}
\begin{equation}\nonumber
  |   \tilde{\delta p}_{\rm s}|  = K \frac{|\mathcal{G}|}{\sqrt{\varepsilon^2\left[(\kappa+1)X^2-1\right]^2+X^2\left(X^2-1\right)^2}}\,,
\end{equation}
\begin{equation}\nonumber
 \Re \{\tilde{\delta p}_{\rm s}\}  = K \frac{\varepsilon\left[(\kappa+1)X^2-1\right] \mathcal{G}_{\rm R} + X(X^2-1)\mathcal{G}_{\rm I}}{{\varepsilon^2\left[(\kappa+1)X^2-1\right]^2+X^2\left(X^2-1\right)^2}}\,,
\end{equation}
\begin{equation}\label{p_func_form_fin}
 \Im \{\tilde{\delta p}_{\rm s}\}  = K \frac{\varepsilon\left[(\kappa+1)X^2-1\right] \mathcal{G}_{\rm I} - X(X^2-1)\mathcal{G}_{\rm R}}{{\varepsilon^2\left[(\kappa+1)X^2-1\right]^2+X^2\left(X^2-1\right)^2}}\,,
\end{equation}
\RVA{where $\mathcal{G}_{\rm R} = \Re\{\mathcal{G}\}$ and $\mathcal{G}_{\rm I} = \Im\{\mathcal{G}\}$. In Figure \ref{pressure_components} we plot a sample of the tidal response in terms of the three components, the modulus, the imaginary, and the real part, centered at the resonance. Notable in the figure is the difference between the resonant amplification obtained for the absolute value of the pressure anomaly, or its real part, compared to its imaginary part. Thus one has to be careful in analyzing the tidal amplification induced by the resonance in terms of the surface pressure anomaly compared to its imaginary part, which goes into the computation of the tidal torque. }

\RVA{It is noteworthy that taking Eq.~\eqref{p_func_form_fin} in the low frequency limit ($X\rightarrow0)$ and setting $\mathcal{G}=1$ we obtain
\begin{equation}
    \Im \{\tilde{\delta p}_{\rm s}\} = K\frac{X}{\varepsilon^2+X^2}\,,
\end{equation}
which is similar to the functional form of the imaginary part of the pressure oscillation given in \cite{ingersoll1978venus}, and obtained empirically using GCMs in \cite{leconte2015asynchronous}. This provides the equivalence between the radiative cooling frequency used in our model and the dissipative frequency used in \cite{leconte2015asynchronous}, as we did in Section \ref{Section_Resonance_Amplitude}}.

\section{Estimates of the absorbed tidal energy}\label{App_Alpha_A}
In this section, we compute estimates of the atmospheric opacity parameter $\alpha_{\rm A}$. As defined in the main text (Section \ref{Sec_thermal_forcing}), $\alpha_{\rm A}$ quantifies the fraction of thermal energy that is absorbed by the atmosphere and is actually available for semidiurnal thermotidal dynamics. To our knowledge, this input tidal power is rarely evaluated in the literature. Here we show that it can be estimated from the distributions of heating rates that are measured in the Earth's atmosphere or computed using GCM simulations. We refer in what follows to \cite{chapman1969atmospheric} and \cite{vichare2013diurnal} who provide such distributions for water vapour (${\rm H_2O}$) and ozone ($\rm O_3$). 

First, we consider the vertical and latitudinal profiles shown by \cite{chapman1969atmospheric}. These profiles do not directly quantify the heating rate $J $, but rather the equivalent temperature variation $\tau_{\rm G}$, which approximately corresponds to the temperature oscillation produced by $J^{m,\sigma}$. The input power is thus defined as \citep[][Eqs.~(143-144)]{chapman1969atmospheric}
\begin{equation}
J = \Re \left\{  \sum_{\sigma} \sum_{m} J^{m,\sigma} \left( x , \theta ,\varphi \right) e^{i \sigma \time } \right\},
\end{equation}
with the coefficients $J^{m,\sigma}$ given by
\begin{equation}
\label{relation_J_tau}
J^{m,\sigma} \left( x , \theta , \varphi \right) = i \sigma C_{\rm p} \tau_{\rm G}^{m}{\sigma} \left( x , \theta , \varphi \right),
\end{equation}
where the heat capacity of the gas per unit mass $C_{\rm p} = \mathcal{R}_{\rm s} / \kappa$. Figure 3.2 of \cite{chapman1969atmospheric} shows the normalised vertical distribution of $\tau_{\rm G}^{2,\sigma}$. Following the authors (their Eq.~163), we set $H = 7.6$~km. The semidiurnal distributions of $\tau_{\rm G}^{2,\sigma}$ for water vapour and ozone can be approximated by the functions 
\begin{align}
\label{Jtempwater}
\tau_{\rm G, H_2O} \left( x ,\theta , \varphi \right)& = T_{{\rm H_2O}} V_{{\rm H_2O}} \left( x \right) \sin\theta e^{i 2 \varphi},  \\
\label{Jtempozone}
\tau_{\rm G, O_3} \left( x ,\theta , \varphi\right)& = T_{\rm O_3} V_{\rm O_3} \left( x \right) \sin^2\theta e^{i  2 \varphi},  
\end{align}
with $T_{{\rm H_2O}} = 0.035$~K and $T_{\rm O_3} =  0.25$~K. As the vertical profile of water vapour shown by Figure 3.2 of \cite{chapman1969atmospheric} is an exponentially decaying function of the altitude, the function $V_{{\rm H_2O}} $ of Eq.\eqref{Jtempwater} is set to 
\begin{equation}
V_{\rm H_2O} \left( x \right) = e^{- k_{\rm H_2O} x},
\end{equation}
where $k_{\rm H_2O} \approx 1/3$. Similarly, the vertical profile of ozone absorption can be approximated by a truncated sine function,
\begin{equation}
V_{\rm O_3} \left( x \right) = \left\{ 
\begin{array}{ll}
K_{\rm O_3} \sin \left( \frac{\pi \left( x - x_1 \right)}{x_2 - x_1} \right), & x_1 \leq x \leq x_2, \\
0 & \mbox{elsewhere},
\end{array}
\right.
\end{equation}
with $K_{\rm O_3} = 1.4$, $x_1 = z_1/ H$, and $x_2 = z_2/H$, the altitudes $z_1$ and~$z_2$ being set to $z_1 = 18$~km and $z_2 = 78$~km. 
At a given location on the sphere ($\theta, \varphi$), the total flux absorbed by the atmosphere for the spherical harmonic $Y_{22}$ is defined as 
\begin{equation}
\delta \mathcal{F} \left( \theta , \varphi \right) = \delta \mathcal{F}_{22} Y_{22} \left( \theta , \varphi\right).
\end{equation}
The efficiency parameter $\alpha_{\rm A}$ of our model is basically the ratio between $\delta \mathcal{F}_{22}$ and the corresponding component of the incident Solar flux, which we retrieve in \ref{quadrupolar_forcing_appendix} (Eq.~\ref{App_eq_quadrupolar_flux})  as 
\begin{equation}
\delta F_{22} = \frac{\sqrt{30 \pi}}{16} F_\star. 
\end{equation}
In order to simplify calculations, we assume that the latitudinal profile of ozone can be approximated by the function $\sin^2 \theta$, similar to the latitudinal profile of water vapour. Under this assumption, we rewrite Eqs.\eqref{Jtempwater} and \eqref{Jtempozone} as 
\begin{align}
\tau_{\rm G, H_2O} \left( x , \theta , \varphi \right)& =  \frac{4}{3} \sqrt{\frac{6 \pi}{5}} T_{\rm H_2O} V_{\rm H_2O} \left( x \right) Y_{22} \left( \theta , \varphi \right),  \\
\tau_{\rm G, O_3} \left( x , \theta , \varphi \right)& =  \frac{4}{3} \sqrt{\frac{6 \pi}{5}} T_{\rm O_3} V_{\rm O_3} \left( x \right) Y_{22} \left( \theta , \varphi \right),  
\end{align}
and the corresponding heating rates as 
\begin{align}
J_{\rm H_2O} \left( x , \theta  , \varphi \right)& =   \frac{4}{3} \sqrt{\frac{6 \pi}{5}}  \sigma C_{\rm p} T_{\rm H_2O} V_{\rm H_2O} \left( x \right) Y_{22} \left( \theta , \varphi \right),  \\
J_{\rm O_3} \left( x , \theta , \varphi \right)& =    \frac{4}{3} \sqrt{\frac{6 \pi}{5}} \sigma C_{\rm p} T_{\rm O_3} V_{\rm O_3} \left( x \right) Y_{22} \left( \theta , \varphi \right),
\end{align}
where the imaginary number appearing in Eq.\eqref{relation_J_tau} has been ignored since it does not affect the amplitude of the thermal forcing{\footnote{\RVA{this step would certainly become invalid if one is after  the phase of the forcing rather than merely its amplitude as is our case here.}}. For a given component, the total input power is obtained by integrating the heating rate over the air column
\begin{equation}
\delta \mathcal{F} = \int_0^{+\infty} J \rho_0H dx =\frac{p_{\rm s}}{g} \int_0^{+\infty}J e^{- x} dx,
\end{equation}
 noting that $\rho_0H = \left( p_0 / g \right)$.
This requires computing the integral of the vertical distribution of the heating rates,
\begin{align}
\label{Awater}
A_{\rm H_2O} &  = \int_0^{+\infty}{V_{\rm H_2O} e^{-x} }dx = \frac{1}{1 + k_{\rm H_2O}}, \\\nonumber
A_{\rm O_3} & = \int_0^{+\infty}{V_{\rm O_3} e^{-x}}dx  \\
&=K_{\rm O_3} \pi   \frac{\left( x_2 - x_1 \right) \left( e^{-x_1}  + e^{-x_2} \right)}{\pi^2 + \left( x^2 - x_1 \right)^2}. 
\end{align}
We end up with $A_{\rm H_2O} = 0.75$ and $A_{\rm O_3} = 4.5 \times 10^{-2}$. Therefore, the total fluxes absorbed by water vapour and ozone are given by
\begin{align}
\delta F_{{\rm H_2O}} \left( \theta  , \varphi \right)& =   \delta \mathcal{F}_{{\rm H_2O}; 22}Y_{22} \left( \theta , \varphi \right),  \\
\delta F_{{\rm O_3}} \left( \theta , \varphi \right)& =  \delta \mathcal{F}_{{\rm O_3} ;22}  Y_{22} \left( \theta , \varphi \right),
\end{align}
with the components
\begin{align}
\delta \mathcal{F}_{{\rm H_2O}; 22} & =    \frac{4}{3} \sqrt{\frac{6 \pi}{5}} \sigma C_{\rm p} \frac{p_{\rm s}}{g} T_{{\rm H_2O}} A_{\rm H_2O} ,  \\
\delta \mathcal{F}_{{\rm O_3} ;22} & =   \frac{4}{3} \sqrt{\frac{6 \pi}{5}} \sigma C_{\rm p} \frac{p_{\rm s}}{g} T_{{\rm O_3}} A_{\rm O_3} .
\end{align}
Using $\sigma=1.45\times10^{-4}~{\rm s^{-1}}$,~$C_{\rm p} = 1004.7~{\rm J~kg^{-1}~K^{-1}}$, $p_{\rm s} = 1013.25~{\rm hPa}$, $g = 9.81~{\rm m~s^{-2}}$, and $F_{\star} = 1366~{\rm W~m^{-2}}$, we obtain $\delta F_{22} = 828.8~{\rm W~m^{-2}} $, and 
\begin{align}\nonumber
& \delta \mathcal{F}_{{\rm H_2O}; 22} \approx 102~{\rm W~m^{-2}}, &&  \frac{\delta \mathcal{F}_{{\rm H_2O}; 22}}{\delta F_{22}} \approx 12.3\%,   \\\nonumber
& \delta \mathcal{F}_{{\rm O_3} ;22}  \approx 44~{\rm W~m^{-2}}, &&  \frac{\delta \mathcal{F}_{{\rm O_3}; 22}}{\delta F_{22}} \approx 5.3\%, \\
& \delta \mathcal{F}_{\rm tot; 22} \approx 146~{\rm W~m^{-2}},  &&  \frac{\delta \mathcal{F}_{{\rm tot}; 22}}{\delta F_{22}} \approx 17.6\%. 
\end{align}
Similar estimates can be computed from the profiles shown by \cite{vichare2013diurnal} (see their Figure 4), where the heating rate is plotted as a function of altitude for the main Hough functions structuring the semi-diurnal atmospheric tide on Earth. Without elucidating again the copious details of the procedure, we obtain for the profiles in \cite{vichare2013diurnal} the following:
\begin{align}\nonumber
& \delta \mathcal{F}_{{\rm H_2O}; 22} \approx 116~{\rm W~m^{-2}}, &&  \frac{\delta \mathcal{F}_{{\rm H_2O}; 22}}{\delta F_{22}} \approx 14.0\%,   \\\nonumber
& \delta \mathcal{F}_{{\rm O_3} ;22}  \approx 34~{\rm W~m^{-2}}, &&  \frac{\delta \mathcal{F}_{{\rm O_3}; 22}}{\delta F_{22}} \approx 4.2\%, \\
& \delta \mathcal{F}_{\rm tot; 22} \approx 150~{\rm W~m^{-2}},  &&  \frac{\delta \mathcal{F}_{{\rm tot}; 22}}{\delta F_{22}} \approx 18.1\%. 
\end{align}
Put together, these computed estimates are consistent and suggest that the efficiency parameter is around $\alpha_{\rm A} {\sim} 17{-}18\%$, which is close to the constraint on $\alpha_{\rm A}$ ($14\%,$ Section \ref{Section_Resonance_Amplitude}) obtained by fitting our model parameters to reproduce the present day semidiurnal surface pressure variations.

\section{Benchmarking our model against GCM simulations}\label{appendix_fitting_PAD19}

The work of \cite{auclair2019generic} presents, to our knowledge, the only attempt on recovering the tidal torque exerted about the spin axis of a thermally forced atmosphere in the high frequency regime; i.e. the regime where the Lamb resonance is encountered, using GCM simulations. Specifically, the mentioned work computed the tidal torque from 3D simulations of atmospheric dynamics using the generic version of the LMDZ GCM \citep{hourdin2006lmdz4} in the case of a nitrogen dominated atmosphere around a dry rocky Venus-like planet.

\begin{figure}[t]
\centering
\includegraphics[width=.48\textwidth]{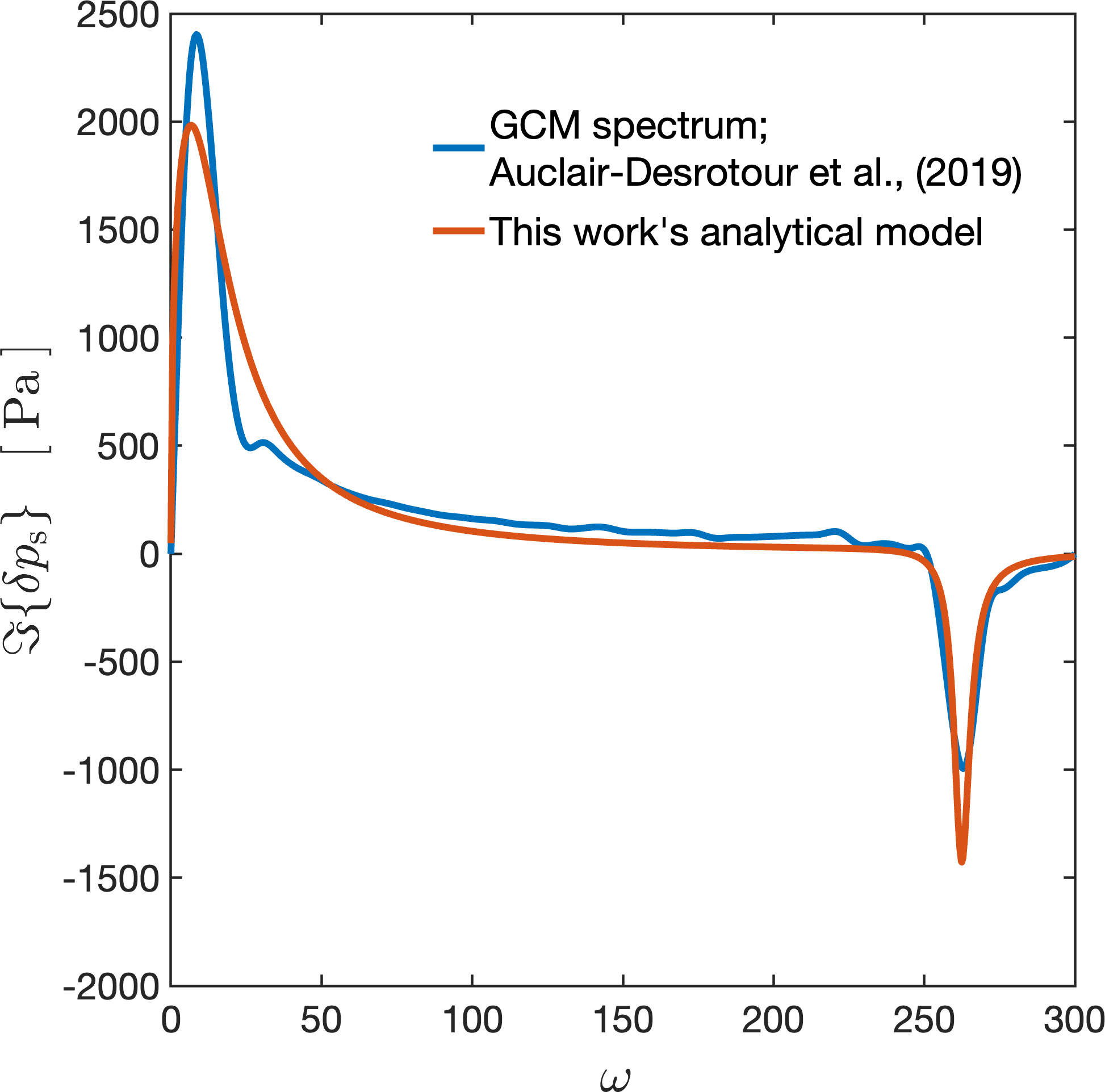}
    \caption{Similar to Figure \ref{Fig_Full_sym_spectrum}, plotted are the tidal spectra, in terms of the imaginary part of the surface pressure anomaly, as a function of the normalized tidal frequency $\omega$. The blue curve is that computed numerically in \cite{auclair2019generic} using 3D GCM simulations, while the orange curve is recovered from our model by tuning the two free frequencies $\sigma_{\rm bl}$ and $\sigma_0$ as discussed in \ref{appendix_fitting_PAD19}. }
    \label{Fig_PAD19_fitted}
\end{figure}

Imposing on our analytical solution (Eq.\ref{imag_pressure_anomaly}) the same planetary parameters defined in Table 1 of \cite{auclair2019generic}, for a 10 bar atmosphere, and setting $\sigma_{\rm bl} = 10^{-5}$ s$^{-1}$ and $\sigma_{\rm 0} = 4\times10^{-6}$ s$^{-1}$, we recover in Figure \ref{Fig_PAD19_fitted} a tidal spectrum that reasonably fits the GCM computed spectrum. In particular, we recover the two amplified responses of the torque in the vicinity of synchronization and at the Lamb resonance, and we recover the asymmetry in the Lamb resonance where the positive peak of the torque is annihilated. Furthermore, we recover a similar scaling of the torque with the forcing frequency $\sigma$. Namely, for the GCM spectrum, the tidal torque $\mathcal{T}\propto\sigma^{0.75}$ near synchronization, and in the high frequency regime  $\mathcal{T}\propto\sigma^{-1}$; while for our model, $\mathcal{T}\propto\sigma^{0.53}$ in the former regime, and $\mathcal{T}\propto\sigma^{-1}$ in the latter.

\section{Atmospheric compositional variations and the possible interplay between shortwave and longwave heating}\label{Appendix_compositional}

\RVA{When studying the case of the Earth in the main text, we have assumed that the atmosphere is always heated by the shortwave. We have thus ignored the asynchronous heating in the infrared, which, as we showed in Section~\ref{Section_Breaking_symmetry}, breaks the symmetry of the Lamb resonance by attenuating the accelerative peak of the tidal torque. The results obtained for the amplitude of the accelerative tidal torque at the resonance can thus be considered as the optimistic limit of the resonant amplitude, as the latter can be attenuated once the longwave heating starts to play a significant role. However, an additional attenuation effect is delivered by the more efficient radiative cooling on a warmer Earth. Namely, the radiative cooling frequency features a strong temperature dependence that scales as $\sigma_0\propto T^3$. Henceforth, the temperature increase required to place the resonance occurrence in the Precambrian would also attenuated the resonant amplitude of the torque by increasing $\sigma_0$. Here we intend to elaborate further on this signature, which is driven by atmospheric compositional variations, by studying the functional form of the tidal response obtained in \ref{App_functional_Form}.}

\RVA{The form of the pressure anomaly in Eqs.~\eqref{p_func_form_fin} allows for an analytical investigation of the Lamb resonance. Namely, we can characterize the positive and negative  peaks of the resonance by studying the derivative of the imaginary part of the pressure anomaly. We focus on the limit studied for the Earth in the main text and set $\mathcal{G}=1$, that is, we assume thermotidal heating is dominated by direct shortwave absorption. Computing the roots of the derivative of Eq.~\eqref{p_func_form_fin} and substituting them in the expression, we obtain the amplitude of the peaks around the resonance. Namely, we denote by $\eta$ the ratio between the value of the tidal torque reached at the peaks around resonance to that at present. Considering the predominant terms of the obtained roots, this amplification ratio reads as
\begin{equation}\label{eta_equation}
    \eta \approx \frac{K_\oslash}{K_\oplus}\frac{\varepsilon_\oplus}{\varepsilon_\oslash}\eta_\oplus\,,
\end{equation}
where we have denoted by the subscript $\oslash$ the parameter values at the resonance, in contrast with the present values denoted by $\oplus$.}

\RVA{Now we allow for variations in $K$ that are induced by the parameters entering into the definition of Eq.~\eqref{Factor_K}. Namely, presuming that $\kappa,R_{\rm p},g,$ and $\Lambda_2$ are indeed constant, the rest of the parameters would modify the expression of the amplification factor $\eta$ (Eq.~\ref{eta_equation}) into:
\begin{equation}
\eta = \eta_\oplus\!\left( \frac{\alpha_{\rm A; \oslash}}{\alpha_{\rm A ; \oplus}} \right)\!\left( \frac{F_{\star ; \rm \oslash}}{F_{\star : \oplus}} \right)\! \left( \frac{p_{\rm s ; \oplus}}{p_{\rm s ; \oslash}} \right) \!\left( \frac{\sigma_{\rm L; \oplus}}{\sigma_{\rm L ; \oslash}} \right)^2\!\!\left(\frac{\sigma_{0;\oplus}}{\sigma_{0;\oslash}}\right) ,
\end{equation}
The resonant frequency in the adiabatic limit scales as $\sigma_{\rm L}~\propto~T_{\rm s}^{1/2}$. Moreover, the radiative cooling frequency is also a function of the surface temperature and it scales as $\sigma_0\propto p_{\rm s}^{-1}T_{\rm s}^{3}$ \citep[e.g., Eq. 10 of ][]{showman2002atmospheric}. Including these scaling relationships in the latter expression of the amplification factor one finds that}
\RVA{\begin{equation}
\eta = \eta_\oplus\left( \frac{\alpha_{\rm A; \oslash}}{\alpha_{\rm A ; \oplus}} \right) \left( \frac{F_{\star ; \rm \oslash}}{F_{\star : \oplus}} \right) \left( \frac{T_{\rm s; \oplus}}{T_{\rm s ; \oslash}} \right)^4.   
\end{equation}
The equilibrium surface temperature varies as $T_{\rm s} \propto \left( \alpha_{\rm h} F_{\star} \right)^{1/4}$, where $\alpha_{\rm h}$ is the fraction of flux that is absorbed by the atmosphere,  mainly in the infrared absorption. It follows that 
\begin{equation}\label{eta_final}
\eta = \eta_\oplus\left( \frac{\alpha_{\rm A; \oslash}}{\alpha_{\rm A ; \oplus}} \right) \left( \frac{\alpha_{\rm h; \oplus}}{\alpha_{\rm h; \oslash}} \right).
\end{equation}
This interesting and simplified version of the amplification factor shows that $\eta$  depends on the opacity of the atmosphere in the visible (via $\alpha_{\rm A})$ and in the infrared (via $\alpha_{\rm h})$, which are functions of the composition. Namely, the two absorption coefficients have opposite effects on $\eta$: a past atmosphere that is  more opaque in the visible features a greater amplification factor, and thus prompts a stronger tidal response upon the resonance occurrence. In contrast, a larger past opacity in the infrared tends to attenuate the resonance amplified response. Thus the bottom line is, increased greenhouse gases in the past (mainly CO$_2$), tend to decrease the amplitude of the accelerative resonant response of the atmosphere via, first, Eq.~\eqref{eta_final}, and via increasing the contribution of the asynchronous thermotidal heating that induces a phase lag on the tidal bulge and consequently attenuates the accelerative torque around the resonance. } 

\section{The resonant period in the isothermal limit}\label{App_isothermal_LOD}
Here we discuss the limit where the temperature profile of the atmosphere is modelled by an isotherm, in contrast with the adiabatic profile we assume in our neutrally stratified model. The isothermal limit  translates to a positive and constant value of the Brunt–Väisälä frequency $N_{\rm B}$, and it has been studied earlier using the linear theory of atmospheric tides \citep[e.g.,][]{lindzen1978effect,lindzen1972lamb,auclair2019generic}. We are specifically interested in estimating the difference between the two models in terms of the Lamb resonance period. Namely, we aim to define the spectral position of the resonance in the isothermal limit and compare with what we obtained for our model in Eq.~\ref{eq_LOD_Ts}.

We adopt the analytical computations in \cite{auclair2019generic} that yield a frequency-dependence for the pressure surface anomaly in the form (their Eq. 12):
\begin{equation}
    \Im\{\delta p_{\rm s}\} = p_{\rm s}\frac{\kappa J_{\rm s}}{gH\sigma}\frac{\frac{H}{h}\left(b_{\rm J}+\frac{1}{2}+\kappa\right)-\frac{1}{2}(b_{\rm J}+1)}{\left[b_{\rm J}(b_{\rm J}+1)+\frac{\kappa H}{h}\right]\left(\frac{H}{h}-\frac{1}{\Gamma_1}\right)}.
\end{equation}
Here the parameters and variables are equivalent in definition to those we use in our work, and $h$ is the equivalent depth. This equation defines the position of the Lamb resonance by the equivalent depth $h_{\rm L}=\Gamma_1 H$ \citep[see also][]{lindzen1972lamb}. Hence, introducing the definition of the equivalent depth
\begin{equation}
    h_n= \frac{R_{\rm p}^2\sigma^2}{\Lambda_n g},
\end{equation}
we obtain, for the fundamental mode, the resonant frequency and consequently the resonant rotational period in the form:
\begin{equation}\label{Eq_LODres_isothermal}
     {\rm LOD_{res;\,iso}}= \frac{4\pi R_{\rm p}}{\sqrt{\mathcal{R_{\rm s}}\Lambda_2 \Gamma_1\overline{T}}+2R_{\rm p}n_\star}.
\end{equation}
This equation differs from that in the neutrally stratified limit (Eq.~\ref{eq_LOD_Ts}) in the fact that temperature here cannot be assumed to be the surface temperature $T_{\rm s}$, and the $\Gamma_1$ factor multiplied by the temperature. To compare the two equations, we assign for the isothermal model a constant value of temperature corresponding to a mass-average of the temperature profile in the neutrally stratified case assuming a fixed total enthalpy of the air column. Namely, using Eq.~\ref{profiles}, we obtain $\overline{T}= T_{\rm s}/(\kappa +1)$, thus we rewrite Eq.~\ref{Eq_LODres_isothermal} as:
\begin{equation}
      {\rm LOD_{res;\,iso}}= \frac{4\pi R_{\rm p}}{\sqrt{\mathcal{R_{\rm s}}\Lambda_n {T_{\rm s}}/{(1-\kappa^2)}}+2R_{\rm p}n_\star}.
\end{equation}
This equation yields a resonant rotational period that is roughly one hour smaller than that obtained in the neutrally stratified case; i.e., the curve shown in Figure \ref{LOD_Ts} shifts vertically downwards by ${\sim}1$~hr.  This places the resonance for present conditions around 21.6 hr, which is closer to the 21.3 hr value obtained by \cite{zahnle1987constant}.
\end{document}